\documentclass[sigplan]{acmart}

\setcopyright{none}
\renewcommand\footnotetextcopyrightpermission[1]{}

\AtBeginDocument{%
  }

\usepackage{hyperref}
\usepackage{algorithm}
\usepackage{algpseudocode}
\usepackage{color}
\usepackage{pifont}
\usepackage[frozencache,cachedir=arxiv-cache]{minted}
\usepackage{xcolor}
\usepackage{ebproof}
\ebproofset{separation=1em}
\usepackage{mathtools}
\usepackage{algorithm}
\usepackage{algpseudocode}
\usepackage{seqsplit}
\usepackage{hyphenat}

\definecolor{serial_blue}{HTML}{DAE8FC}
\definecolor{dev1green}{HTML}{D5E8D4}
\definecolor{dev2yellow}{HTML}{FFF2CC}

\setminted{
    fontsize=\tiny,
    breaklines,
    xleftmargin=0em,
    numbersep=6pt
}

\usepackage{subcaption}
\ifthenelse{0=1}{
\newcommand{\TODO}[1]{\textcolor{red}{TODO: #1}}
\newcommand\help[1]{\textcolor{green}{Help Needed: #1}}
\newcommand\willow[1]{\textcolor{olive}{Willow: #1}}
\newcommand\adrian[1]{\textcolor{red}{Adrian: #1}}
\newcommand\sangyoon[1]{\textcolor{purple}{Sang Yoon: #1}}

}{
\newcommand{\TODO}[1]{}
\newcommand\help[1]{}
\newcommand\willow[1]{}
\newcommand\adrian[1]{}
\newcommand\sangyoon[1]{}

}

\begin{document}

\title{WingSpan: Concurrency and Dependence for Sparse and Structured Tensor Compilers}

\author{Adrian Gushin}
\affiliation{%
  \institution{Georgia Institute of Technology}
  \city{Atlanta}
  \state{Georgia}
  \country{USA}}
\email{agushin3@gatech.edu}

\author{Sang Yoon Kim}
\affiliation{%
  \institution{Georgia Institute of Technology}
  \city{Atlanta}
  \state{Georgia}
  \country{USA}
}
\email{sangyoon@gatech.edu}

\author{Willow Ahrens}
\affiliation{%
 \institution{Georgia Institute of Technology}
 \city{Atlanta}
 \state{Georgia}
 \country{USA}}
\email{ahrens@gatech.edu}

\renewcommand{\shortauthors}{Gushin et al.}

\begin{abstract}
Sparse tensors represent data that is mostly zero or some other compressible fill pattern. Such datasets can be massive, so optimized tensor algebra libraries and compilers have been developed to exploit these patterns to improve performance. Existing systems, however, frequently lack support for parallelism, especially when outputs are sparse or multiple inputs are sparse. We propose WingSpan, a sparse tensor language enabling unrestricted parallel programming. WingSpan supports arbitrary composition of parallel loops and data structures, matching or exceeding the performance of hand-optimized parallel routines on critical kernels such as SpGEMM. We also introduce a dependence theory for the safety of parallel programs involving sparse tensors and structures beyond sparsity.
\end{abstract}

\keywords{Sparse Tensors, Parallel, Programming Languages}

\maketitle

\makeatletter
\fancyhead[L]{\@headfootfont\shorttitle}
\fancyhead[R]{\@headfootfont\shortauthors}
\makeatother

\section{Introduction}

Datasets in the modern era have become massive, but many tensors contain patterns allowing for compressed representations that make computation tractable.
Sparse tensors contain mostly zero or ``fill'' values; additional patterns, such as non-zero blocks or identical-value runs, enable further compression \cite{donenfeld_unified_2022}. Sparsity occurs naturally in real-world applications with community structure, such as graph algorithms \cite{kepner_graph_2011}, quantum simulations \cite{venev_qblaze_2025}, or neural networks \cite{zhu_survey_2024,frankle_lottery_2019}. Efficient sparse kernels are essential to handle these applications at scale \cite{de2019deep, hochbaum2016sparse, el2013understanding, jenkinson2014machine}.

Optimizing sparse applications can be complex. Due to algebraic properties like $x * 0 = 0$, only nonfill values must be stored and processed. However, operating directly on compressed data structures involves intricate traversals and invariants. To address this complexity, a variety of compiler approaches have emerged to automatically generate efficient kernels. Frameworks such as TACO, Finch, and SparseTIR support a wide range of data structures, hardware architectures, and optimizations ~\cite{kjolstad_taco_2017, ahrens_finch_2025, chou_format_2018, ye_sparsetir_2023, bik_compiler_2022}.

Among parallel targets, multicore CPUs are critical for sparse compilers. CPU-exclusive supercomputers like Fugaku compete for first place globally when evaluated on sparse benchmarks like the Graph500 \cite{graph500}. CPU performance is key to enabling compute-intensive tasks, like mobile computer vision, on edge and embedded devices \cite{kong2025simulating}. Finally, sparse kernels like SpGEMM involve extensive control flow, making CPUs a natural target for their highly optimized branch predictors and management logic \cite{won2026insum}.

Unfortunately, existing sparse tensor compiler support for parallelism is limited and fragmented. TACO, for example, only supports parallelism when the output is dense, and can only load balance effectively when only one input is sparse \cite{senanayake_sparse_2020}. MLIR Sparse supports sparse outputs by creating a large temporary buffer full of each individual summand together with its coordinate, then sorts and sums them later \cite{bik_compiler_2022}. Other compilers, such as Taichi or SparseTIR, have similar constraints \cite{hu_taichi_2019,ye_sparsetir_2023}. All of these systems require intricate manual tuning to achieve good performance.

There are several challenges in parallel sparse algorithms:

\noindent\textbf{Complex Data Structure Invariants:} Sparse data structures are complex, interdependent, and difficult to update concurrently. Even the most common formats, such as CSR, involve prefix sums and large contiguous data buffers. Taking Figure \ref{fig:structuraldiversity} as an example, several formats have complex nested structures that require careful synchronization to update safely in parallel. Algorithms may parallelize along different data structure dimensions with different dependence guarantees. Sparse parallel programming paradigms must therefore safely access these complex structures without breaking critical invariants. Doing so while maintaining scalability across multiple cores is a central challenge.

\noindent\textbf{Irregular Compute Balance:} Sparse workloads are often imbalanced. When a single input is sparse, the non-zeroes can be treated as proportional to the work \cite{senanayake_sparse_2020}. However, when multiple inputs are sparse, the amount of work per non-zero is not easily statically computable and can vary significantly at runtime \cite{dasgupta_sparse_2012}. Dynamic load balancing would be an effective strategy to scale with near-optimal efficiency, but computing dynamic pieces of a sparse output directly is a challenge, and existing compilers do not support it.

\noindent\textbf{Weak Theoretical Foundation:} Dependence analysis verifies the safety of compiler optimizations in dense code, such as tiling and parallelization, by assuming that distinct coordinate tuples imply isolated memory locations \cite{kennedy2001optimizing}. For example, threads modifying $A[1,1]$ and $A[2,1]$ cannot race because $(1,1) \neq (2,1)$. This assumption fails in sparse formats, where distinct indices may share indirection arrays or depend on cumulative sums. New theoretical tools are necessary to safely parallelize sparse tensor programs.

\textbf{To address these challenges, we present WingSpan, a high-level parallel array programming language} for sparse tensor algebra that compiles to performant LLVM code. Existing sparse tensor compilers such as TACO or Finch can describe sparse and structured data hierarchically, associating each dimension with a level in a tree annotated with format descriptors. These descriptors denote the level's data representation, such as dense or sparse. Finch is notable for supporting a surprising variety of formats, including block and run-length encoded levels. WingSpan builds on Finch to make the following three contributions:

\textbf{1. Concurrent annotations for Finch's hierarchical sparse tensor format description language.} We add the mutex, shard, and merge level wrappers, diagrammed in Figure \ref{fig:parallel-levels}. We build the first compiler that supports parallel sparse outputs directly, without intermediates. WingSpan's new formats extend and interoperate with the entire existing format language to facilitate parallelism.

\textbf{2. A sparsity-aware parallel programming language}, supporting nested parallelism, parallelism around and within sparse workspaces, and static and dynamic scheduling. Ours is the first language to support arbitrary composition of parallel loops, and the first to support arbitrary interactions between parallel loops and sparse workspaces. Our abstraction comes at no cost to performance; we match or exceed state-of-the-art parallel systems on tensor applications like SpGEMM ($1.21\times$ geomean speedup), SpAdd ($4.22\times$), Hadamard product ($3.39\times$), SpMSpV ($1.47\times$), and MTTKRP ($0.96\times$).

\textbf{3. The first dependence theory of parallel programming safety for hierarchical sparse and structured tensor formats}. Our theory accounts for all of the dense, sparse, and otherwise structured tensor descriptions in Finch's language, which is to our knowledge the largest set of such formats \cite{ahrens_finch_2025}. Our new theoretical foundation enables developers to prove their algorithm parallelizes without race conditions regardless of the tensor types used in the program.

\section{Background}

\subsection{Fiber Trees} \label{sec:ft}

Fiber-tree style tensor abstractions for sparse and structured data are abundant
\cite{sze_efficient_2020,chou_compilation_2022,chou_format_2018}.  
Fiber-trees represent a multi-dimensional tensor as a nested vector structure, where each level of the nesting corresponds to a dimension of the
tensor. For example, a matrix would be represented as a vector of vectors. Fiber-trees can represent sparse tensors by varying the format of
vectors used at each level in a tree. Thus, a CSC sparse matrix might be represented
as a dense vector of sparse vectors. Nodes in the tree are referred to as \textbf{fibers}.

\begin{figure}
\centering
  \includegraphics[width=\linewidth]{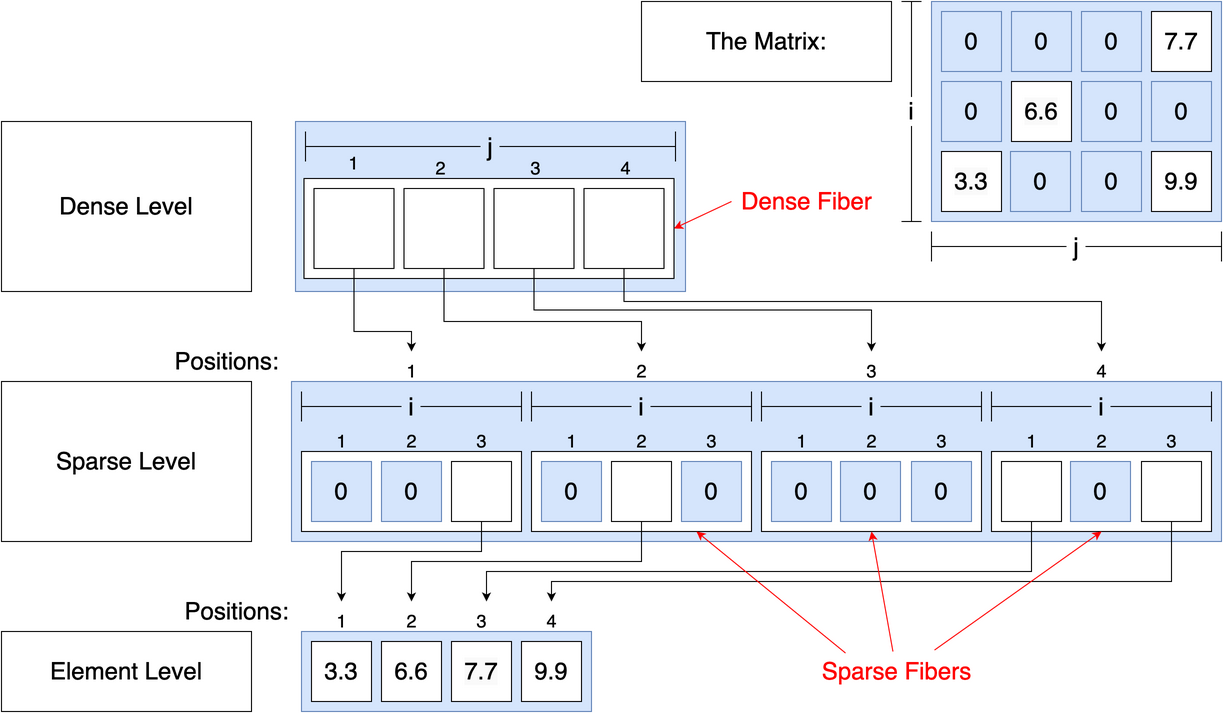}
  \caption{A fiber tree representation of a sparse matrix in CSC format, with a dense outer level, a sparse inner level, and an element level of leaves that store values \cite{ahrens_finch_2025, sze_efficient_2020}. Note that each index within a level's fibers maps to a position in its child level. When the child is a leaf, this map materializes directly to a payload (value).}
  \label{fig:levelsvsfibers}
\end{figure}

Instead of storing the data for each subfiber separately, most sparse tensor formats, such as CSR, DCSR, and COO, usually store the data for all fibers in a level contiguously. In this way, we can think of a level as a bulk allocator for
fibers. Continuing the analogy, each fiber is
disambiguated by a \textbf{position}, or an identifier in the bulk pool of
subfibers. When we need to refer to a particular fiber $f$ at position $p$ in level $l$ of tensor $T$, we may write $f = \texttt{fiber}(T, l, p)$. Figure~\ref{fig:levelsvsfibers} shows a simple example of a level as a pool of fibers.
The children of fibers can be addressed with indices, expressible as $f(i)$. If the child is a subfiber at position $q$, then $f(i) = \texttt{child}(q)$. If the child is a constant $z$, such as zero, then $f(i) = \texttt{payload}(z)$.

\subsection{Finch Tensor Compiler}

\begin{figure}[ht]
    \centering
    \includegraphics[width=0.9\linewidth]{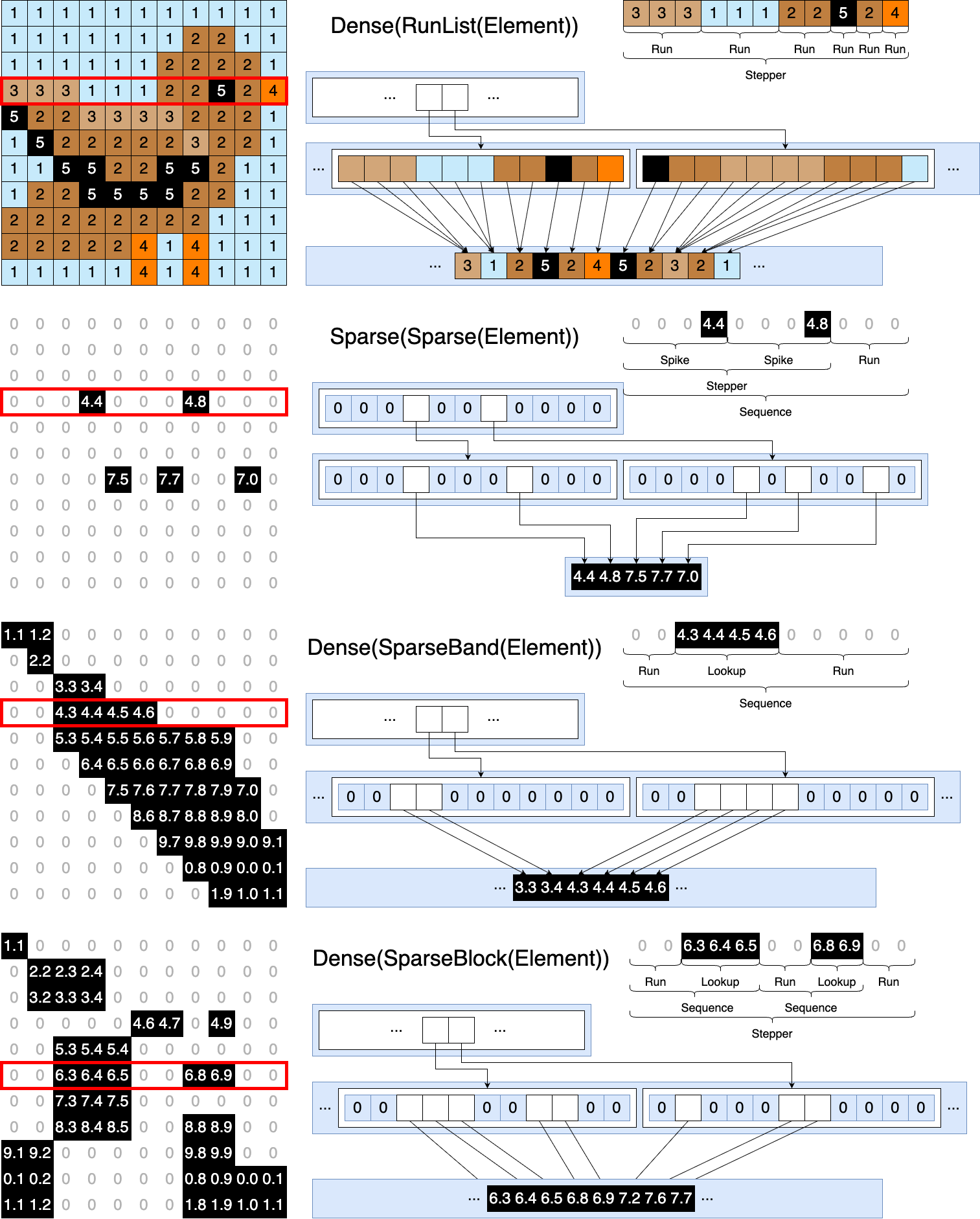}\hfill%
    \caption{Examples of sparse matrix structures represented in the hierarchical format language of Finch\cite{ahrens_finch_2025}.}
    \label{fig:structuraldiversity}
\end{figure}

Finch, a state-of-the-art tensor compiler, is unique for its great diversity of level formats \cite{ahrens_finch_2025}. It extends level-by-level fibertree descriptions to capture any combination of banded, triangular, run-length-encoded, padding, one-hot-encoding, or sparse datasets, as shown in Figure~\ref{fig:structuraldiversity}.

WingSpan uses the Finch compiler to support lowering to LLVM code. Particularly, it exploits Finch's hierarchical tensor decomposition to integrate the \textit{modifier} levels shown in Figure \ref{fig:parallel-levels}. We refer to WingSpan's levels as modifiers to emphasize that, unlike standard levels, they are dimensionless. Rather, they construct the data structures and code lowering algorithms necessary to parallelize Finch code.

WingSpan's strategy is sufficiently general that it could be implemented using any state-of-the-art tensor compiler with hierarchical formats, such as TACO. We use Finch because it supports the greatest diversity in physical level formats. WingSpan's modifier levels interoperate with any permutation of child levels, so Finch's greater format diversity maximizes the language's expressive power. This flexibility enables WingSpan to parallelize a larger number of kernels across a larger combination of tensor storage choices. 

Because WingSpan builds atop of Finch, our code is integrated into the larger Finch compiler. The entire system is available online at the Finch GitHub repository. \footnote{"Wingspan is merged to main in the Finch repository, available online at https://github.com/finch-tensor/Finch.jl.} We also provide our harness for our performance section. \footnote{Our benchmark harness is available online at https://github.com/finch-tensor/FinchBenchmarks/tree/cgo26-artifact.}

\subsection{Classical Dependence Analysis}

Safe parallel code requires avoiding race conditions. Classical dependence analysis can determine when it is possible to safely transform or parallelize loops over dense arrays\cite{kennedy2001optimizing}.

To introduce vocabulary, consider an \textbf{$n$-dimensional tensor} $A$ mapping $n$-tuples of integer \textbf{coordinates} to values. The \textbf{access} expression $A_{ijk\cdots}$ refers to the value stored at a particular location $ijk...$. The terms $i,j,k...$ may be expressions and are referred to as \textbf{tensor indices} or \textbf{subscripts}.
We refer to the loop iteration variable as a \textbf{loop index}. For example, in the loop ``\texttt{for i in 1:m}'', $i$ is the loop index. 
In the style of Fortran, we index from 1 and assume column-major order, so the first subscript of a tensor has unit stride.

We say that two accesses are \textbf{dependent} if one of them is a write and they access the same memory location, and \textbf{independent} otherwise.
Dependence extends to multidimensional tensors by requiring that all subscripts match pairwise across the two accesses.
A dependence may occur between an access expression and itself in different loop iterations, or between two accesses in the same iteration. Independence requires that no two loop iterations access the vector at the same subscript value, which implies that there is no distribution of iterations to threads that could cause a data race. Figure \ref{fig:indep-dep} gives a simple example.

\begin{figure}[t]
\centering

\begin{subfigure}[t]{0.5\columnwidth}
\begin{minted}{julia}
for i=parallel(_)
    c[i] += 1
end
\end{minted}
\label{fig:indep-dep-indep}
\end{subfigure}%
\hfill
\begin{subfigure}[t]{0.5\columnwidth}
\begin{minted}{julia}
for i=parallel(_)
    c[i - i % 2] += 1
end
\end{minted}
\label{fig:indep-dep-dep}
\end{subfigure}
\caption{Two example kernels. The access is independent on the left because no two loop iterations will modify $c$ with the same value for $i$. Thus, the kernel is safe to parallelize. The access on the right is dependent on itself since adjacent loop iterations modify the same element of $c$. This potential race means that the kernel may not be safe to parallelize.}
\label{fig:indep-dep}
\Description{Two different kernels accessing a vector c. The first iterates over all of c's values in parallel, with each iteration incrementing c at index i by 1. The second is similar, but pairs up adjacent iterations by accessing c at i minus i mod 2. The first kernel is independent because no two iterations of the loop will access c using the same value for c. In the second, i is dependent because adjacent loop iterations access c using the same value}
\end{figure}

\section{Level Dependence Testing} \label{sec:atom-concurr}

\begin{figure}[b]
    \centering
    \includegraphics[width=0.9\linewidth]{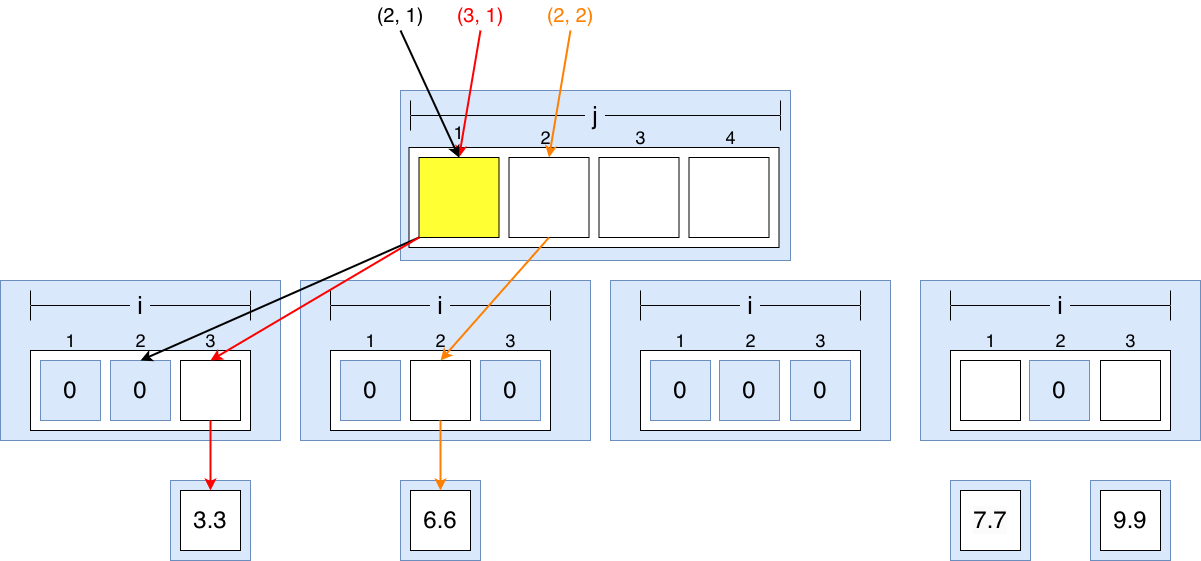}
    \caption{A fibertree disambiguation of a column-major $3\times4$ sparse tensor. Three memory accesses are being made to subscripts $(2, 1)$, $(3, 1)$, and $(2, 2)$, denoted by black, red, and orange arrows respectively. Observe that despite all the subscripts referring to different elements, pairs $(2, 1)$ and $(3, 1)$ access the same memory location in the first level of the tensor (highlighted in yellow). In contrast, subscript (2, 2) accesses unique memory locations across the entire fibertree.}
    \label{fig:level_dep}
\end{figure}

The foundational assumption undergirding classical dependence analysis is that distinct subscript tuples refer to distinct memory locations. This assumption holds for dense tensors because they map directly to contiguous memory regions. However, unlike dense arrays, sparse arrays are stored in tree structures built from multiple levels. Materializing a single coordinate results in \textit{multiple} memory operations across different levels of the tensor tree. Figure \ref{fig:level_dep} shows that seemingly independent accesses can result in collisions in fibertrees. Reasoning about dependence in the sparse case therefore requires examining each level of the tree separately.

\subsection{Dependencies on Tensor Trees}

Classical dependence can be generalized to tree structures through \textbf{node dependence}. Two subscript tuples are \textbf{level-$k$ node dependent} on tensor A if they access the same memory location on level $k$ of the fibertree. The language of classical theory is sufficient to capture this nuance. For a level $k$ in a leaf-to-root path $k, \cdots, q, p$, the subscript accessing $k$ is $(k, \cdots, q, p)$. For example, in Figure \ref{fig:level_dep}, level 1 is accessed by subscript $j$ and level 2 by $(i, j)$. Using a dependence test like Banerjee's inequality on level $k$'s subscript can detect a level-$k$ node dependence \cite{banerjee1997dependence}. Therefore, using these tests on every level's access subscript will compute the set of all node dependencies on the tree. In Figure \ref{fig:level_dep}, testing subscript $j$ would detect the level-1 node dependence from the race on $j=1$, and disprove level-2 node dependence because its subscript $(i,j)$ is always accessed with unique pairs.

While node dependence begins to generalize the language of dependence analysis to tensor tree representations, it is not complete. Sparse and structured levels are often composed of complex data buffers with invariants that can be violated even if parallel accesses occur at different nodes. For instance, the compressed sparse column format uses several buffers maintaining cumulative prefix sums that must remain monotonically non-decreasing \cite{chou_format_2018}. Otherwise safe parallel updates to these arrays could induce data races by inadvertently violating these invariants.

The presented sparse representations are merely a single problem case. There are myriad other tensor format descriptors, each of which has idiosyncratic invariants. Writing a separate dependency theory for each one is not feasible. We generalize by analyzing the three types of level dependencies parallel accesses can create. To introduce these definitions, recall from Section \ref{sec:ft} that levels are pools of fibers, and the fiber at position $p$ in level $l$ in tensor $T$ is denoted as $f = \texttt{fiber}(T, l, p)$ and can be accessed at index $i$ as $f(i)$.

Under this construction, a parallel dependence between two fiber accesses $\texttt{fiber}(T, k, p)(i)$ and $\texttt{fiber}(T, k, p')(i')$ in level $k$ can be categorized into three types:
\begin{itemize}
    \item \textbf{Level-$k$ node dependence}
    if the accesses modify the same fiber and the same index, $p' = p$ and $i' = i$.
    \item \textbf{Level-$k$ sibling dependence} if the accesses modify the same fiber with different indices, $p = p'$ and $i \neq i'$.
    \item \textbf{Level-$k$ cousin dependence}
    if the accesses are made to different fibers, $p \neq p'$.
\end{itemize}

This vocabulary allows for greater precision when discussing level dependencies. For example, in Figure \ref{fig:level_dep}, the black and red accesses to $(2, 1)$ and $(3, 1)$ create a level-1 node dependence and a level-2 sibling dependence. Similarly, the red and orange accesses to $(3, 1)$ and $(2, 2)$ produce a level-1 sibling and a level-2 cousin dependence. 

Different levels have different representations, some of which can safely support these access patterns. To track these safety guarantees while maintaining generality, each tensor has a \textbf{stability vector} $S$. For an $n$-mode tensor, $S$ is a vector of $n+1$ sets. $S_1$ represents the tensor's leaf (element) level, and every subsequent entry corresponds to a mode in the tensor. For a mode $k$, $S_{k+1}$ is the subset of $\{$node, sibling, cousin$\}$ dependencies that the level representing $k$'s dimension can support without causing undefined behavior.

\begin{algorithm}[b]
    \begin{algorithmic}
        \Require $A$ \Comment{The tensor to be accessed}
        \Require $S$ \Comment{A's stability vector.}
        \Require $p$ \Comment{The loop we wish to parallelize}
        \Require $f_1(i...), ..., f_m(i...)$ \Comment{Subscripts of first access}
        \Require $g_1(i'...), ..., g_m(i'...)$ \Comment{Subscripts of second access}
        \Require $i...$ the indices of loops within the scope of $A$'s declaration which contain both accesses, from root to leaf.
        \For{$j = 1, ..., m+1$}
            \State $d \gets \mathop{dep\_test}((f_{j}(i...), ..., f_{m}(i...)),$
            \Statex $\hspace{4.5em}(g_{j}(i'...), ..., g_{m}(i'...)))$
            \State $race_{position} = d_1, ..., d_{p-1} \in \{=,*\} \text{ and } d_p \not\in\{=\}$
            \If{$j > 1$}
                \State $d' \gets \mathop{dep\_test}((f_{j-1}(i...), ..., f_{m}(i...)),$
                \Statex $\hspace{4.5em}(g_{j-1}(i'...), ..., g_{m}(i'...)))$
                \State $race_{index} = d'_1, ..., d'_{p-1} \in \{=,*\} \text{ and } d'_p \not\in\{=\}$
            \EndIf
            \State $W_j \gets \varnothing$
            \If{$i_p \not\in f_j,... f_m, g_j, ..., g_m$ and $i_p \in f_1...f_m, g_1,...g_m$}
                \State{\textbf{continue}} \Comment{Skip levels not accessed in parallel}
            \EndIf
            \If {not $race_{position}$}
                \State $W_j \gets W_j \cup \text{\{cousin\}}$
            \EndIf
            \If {$race_{position}$ and not $race_{index}$}
                \State $W_j \gets W_j \cup \text{\{sibling, cousin\}}$
            \EndIf
            \If {$race_{index}$}
                \State $W_j \gets W_j \cup \text{\{node, sibling, cousin\}}$
            \EndIf
        \EndFor
        \State \Return $W$
    \end{algorithmic}
    \caption{A dependence test for a single pair of accesses. The algorithm returns $W$, the required stability for each level of the tensor. Note that $race_{position}$ is true when there is a possible race on the position of the level, and $race_{index}$ is true when there is a possible race on both the position and index. The access pair is safe if $W_j \subseteq S_j$ for all $j$.}
    \label{alg:level_dep_test}
\end{algorithm}

The lack of stability in Finch's existing level formats reveals the difficulty in sparse parallelism. Element (leaf) levels have no children, but cannot support data races on the leaves, so $S_1 = \{$sibling, cousin$\}$. Additionally, if $m$'s level is dense, $S_{m+1} = \{$node, sibling, cousin$\}$ because dense levels are static mapping functions. Unfortunately, for any other level format, $S_{m+1}=\varnothing$. WingSpan enables sparse and structured parallelism by addressing this gap.

\subsection{The Level Dependence Test}

Dependence tests can detect data races on a dense multidimensional array. In this section, we expand such tests to handle trees. Let $i = (i_1, ... , i_n)$ be the vector of values of loop indices in a loop nest. A standard \textbf{dependence test} determines whether (and when) two accesses $A_{f_1(i...), ..., f_m(i...)}$ in iteration $i$ and $A_{g_1(i'...), ..., g_m(i'...)}$ in iteration $i'$ access the same memory. The result is a \textbf{direction vector} $d$ representing the constraints $i_k d_k i'_k$, where $d_k \in \{<, =, >, *\}$. The $*$ value indicates there is no provable constraint on $i_k$ and $i'_k$.

A loop is safe to parallelize if it does not carry a dependence. Loop $p$ carries a dependence when $d_1...d_p-1$ are in $\{*, =\}$ and $d_p$ is not $=$. If these conditions are true, the parallel loop generates repeated subscripts across multiple iterations. Consequently, different iterations of the same parallel loop might access the same array element and a race condition is possible.  We conservatively include $*$ in the check on $d_1...d_{p-1}$ because $*$ does not disprove a $=$ relationship.

Algorithm \ref{alg:level_dep_test} builds on this notion to determine when the tensor's stability is sufficient for a particular parallel loop. The algorithm computes a loop nest's direction vector and uses the properties of loop carried dependencies to determine the types of parallel accesses a program will make to a tensor. Running Algorithm \ref{alg:level_dep_test} against all subscript pairs and parallel loops in a program evaluates whether a tensor's stability vector is sufficient to prevent race conditions in the program. 

\section{The WingSpan Language}

\newcommand{\hlite}[1]{\colorbox{dev1green}{#1}}
\newcommand{\mhlite}[1]{\mathchoice
    {\colorbox{dev1green}{$\displaystyle#1$}}
    {\colorbox{dev1green}{$\textstyle#1$}}
    {\colorbox{dev1green}{$\scriptstyle#1$}}
    {\colorbox{dev1green}{$\scriptscriptstyle#1$}}}
\begin{figure}[h]
\centering
\tiny
\begin{align*}
    T \in \text{Tensor}&
        \quad l,p,a,b,n,c \in \mathbb{Z}
        \quad v \in \text{Literal}
        \quad i \in \text{Index} \quad \sigma \in \text{Store}\\
      x,y \in &\text{Variable}
        \quad t \in \mhlite{\text{Task}}
        \quad D \in \mhlite{\text{Device}}
        \quad h \in \mhlite{\text{Schedule}}\\
C \in \text{Context} &::= \langle s, t, \sigma \rangle \quad
    m \in \text{Mode} ::= \begin{aligned}[t]
        &\texttt{read}
        \mid \texttt{update}(\oplus)
        \mid \texttt{undeclared}
    \end{aligned}\\
    A \in \text{Access} &::= \texttt{access}(F, i\ldots)\quad
    F \in \text{Fiber} ::= \texttt{fiber}(T, l, p) \\
    e \in \text{Expression} &::= \begin{aligned}[t]
        &v
        \mid \texttt{call}(\otimes, e_1, \ldots, e_n)
        \mid \texttt{access}(F, i\ldots)
    \end{aligned}\\
    r \in \text{Range} &::= a\texttt{:} b \mid \texttt{looplet}(a\texttt{:} b) \mid \mhlite{\texttt{parallel}(r, D, h)}\\
    s \in \text{Statement} &::= 
        s_1 ; s_2 \mid \texttt{if}(e, s)
        \mid \mhlite{s_1 \| s_2}
        \mid C
        \mid x\texttt{=} e
        \mid \texttt{assign}(A, \oplus, e)
        \mid \texttt{skip}\\
        &\mid \texttt{unwrap}(F)
        \mid \texttt{increment}(F, \oplus, e)
        \mid \texttt{unfurl}(F)\\
        &\mid \texttt{declare}(T, z, \oplus, n\ldots)
        \mid \texttt{freeze}(T, \oplus)
        \mid \texttt{thaw}(T, \oplus)\\
        &\mid \mhlite{\texttt{spawn}(s, D, e)}
        \mid \mhlite{\texttt{transfer}(T, s)}
        \mid \texttt{for}\ i = r\ \texttt{do}\ s\ \texttt{end}\\
        &\mid \mhlite{\texttt{worker}(i, r, h, s)}
        \mid \mhlite{\texttt{checkout}(T, l, m, t)}
        \mid \mhlite{\texttt{commit}(T, l, m, t)}
\end{align*}
\caption{The Grammar of WingSpan.  Highlighted terms represent the new concepts added in this paper.}
\label{fig:syntax}
\end{figure}

WingSpan builds on Finch by adding several new language features to facilitate parallelism. WingSpan introduces four modifier levels to support parallelism, the \textbf{mutex}, \textbf{isolate}, \textbf{shard} and \textbf{merge} levels, which wrap Finch levels to protect them. WingSpan coordinates these levels with Finch's existing framework by adding parallel \textbf{\texttt{for}} loops with \textbf{\texttt{parallel}} domains to the language's AST, along with schedulers that specify load balancing strategies, a hierarchy of \textbf{device} and \textbf{task} abstractions, and new \textbf{\texttt{checkout}} and \textbf{\texttt{commit}} interface functions. This section introduces the WingSpan language with syntax (Figure \ref{fig:syntax}) and semantics (Figure \ref{fig:semantics}), which we will reference as we describe the features in more detail.

\newcommand{\rulesep}{\vspace{3pt}}
\begin{figure}[H]
    \centering
    \tiny

    \begin{prooftree}
        \hypo{\forall_i, \langle e_i, t, \sigma \rangle \Rightarrow e'_i}
        \hypo{\langle f, t, \sigma \rangle \Rightarrow g}
        \infer2[$Call$]{\langle \texttt{call}(f, e...), t, \sigma\rangle \rightarrow g(e'...)}
    \end{prooftree}
    \hfill
    \begin{prooftree}
    \hypo{\langle s_1, t, \sigma \rangle \rightarrow \sigma'}
    \infer1[$Block$]{\langle s_1;s_2, t, \sigma \rangle \rightarrow \langle  s_2, t, \sigma' \rangle }
    \end{prooftree}

    \rulesep

    \begin{prooftree}
        \hypo{\langle e, t, \sigma \rangle \Rightarrow true}
        \hypo{}
        \infer2[$IfTrue$]{\langle \texttt{if}(e, s), t, \sigma \rangle \rightarrow \langle \langle s, t, \{\} \rangle, t, \sigma \rangle}
    \end{prooftree}%
    \hfill
    \begin{prooftree}
        \hypo{\langle e, t, \sigma \rangle \Rightarrow false}
        \hypo{}
        \infer2[$IfFalse$]{\langle \texttt{if}(e, s), t, \sigma \rangle \rightarrow \sigma}
    \end{prooftree}%

    \rulesep

    \begin{prooftree}
      \hypo{\langle \texttt{mode}(F), t, \sigma \rangle \rightarrow \texttt{read}}
      \hypo{\langle \texttt{unwrap}(F), t, \sigma \rangle \rightarrow v}
      \infer2[$Access$]{\langle s, t, \sigma\rangle \rightarrow \langle s[\texttt{access}(F) \rightarrow v], t, \sigma\rangle} 
    \end{prooftree}

    \rulesep

    \begin{prooftree}
      \hypo{\langle \texttt{mode}(F), t, \sigma \rangle \rightarrow \texttt{update}(\oplus)}
      \hypo{\langle \texttt{increment}(F, \oplus, e), t, \sigma \rangle \rightarrow \sigma'}
      \infer2[$Assign$]{\langle \texttt{assign}(\texttt{access}(F), \oplus, e), t, \sigma \rangle \rightarrow \sigma'}
    \end{prooftree}

    \rulesep

    \begin{prooftree}
        \hypo{\langle \texttt{unfurl}(F_1); ..., \texttt{unfurl}(F_m); t, \sigma \rangle \rightarrow \sigma'}
        \hypo{\forall_k, \texttt{access}(F_k, j..., i) \in s}
        \infer2[$Unfurl$]{\langle \texttt{for}\ i = a\texttt{:} b\ \texttt{do}\ s\ \texttt{end}, t, \sigma\rangle \rightarrow \langle \texttt{for}\ i = \texttt{looplet}(a\texttt{:} b)\ \texttt{do}\ s\ \texttt{end}, t, \sigma'\rangle}
    \end{prooftree}

    \rulesep
    
    \begin{prooftree}
        \hypo{a \leq b}
        \hypo{\langle \langle s, t, \{i \mapsto a\} \rangle, t, \sigma \rangle \rightarrow \sigma'}
        \infer2[$SerialFor$]{\splitfrac{\langle \texttt{for}\ i = \texttt{looplet}(a\texttt{:} b)\ \texttt{do}\ s\ \texttt{end}, t, \sigma\rangle \rightarrow}{\langle \texttt{for}\ i = \texttt{looplet}(a+1\texttt{:} b)\ \texttt{do}\ s\ \texttt{end}, t, \sigma'\rangle}}
    \end{prooftree}

    \rulesep

    \begin{prooftree}
        \hypo{i' = \sigma[i]}
        \hypo{\texttt{fiber}(T, l, p)(i') = \texttt{payload}(v)}
        \infer2[$Payload$]{\langle s, t, \sigma \rangle \rightarrow \langle s[\texttt{access}(\texttt{fiber}(T, l, p), j..., i) \rightarrow v], t, \sigma \rangle}
    \end{prooftree}

    \rulesep

    \begin{prooftree}
        \hypo{i' = \sigma[i]}
        \hypo{\texttt{fiber}(T, l, p)(i') = \texttt{child}(q)}
        \infer2[$Child$]{\splitfrac{\langle s, t, \sigma \rangle \rightarrow}{\langle s[\texttt{access}(\texttt{fiber}(T, l, p), j..., i) \rightarrow \texttt{access}(\texttt{fiber}(T, l-1, q), j...)], t, \sigma \rangle}}
    \end{prooftree}

    \rulesep
    
    \begin{prooftree}
    \hypo{\langle \texttt{mode}(T), t, \sigma \rangle \rightarrow \texttt{undeclared}}
    \hypo{\langle \texttt{mode}(T), t, \sigma' \rangle \rightarrow \texttt{update}(\oplus)}
    \infer2[$Declare$]{\langle T \texttt{ = declare}(T, z, \oplus, n...), t, \sigma \rangle \rightarrow \sigma'}
    \end{prooftree}

    \rulesep

    \begin{prooftree}
    \hypo{T \in dom(\sigma)}
    \hypo{\langle \texttt{mode}(T), t, \sigma \rangle \rightarrow \texttt{update}(\oplus)}
    \hypo{\langle \texttt{mode}(T), t, \sigma' \rangle \rightarrow \texttt{read}}
    \infer3[$Freeze$]{\langle T \texttt{ = freeze}(T, \oplus), t, \sigma \rangle \rightarrow \sigma'}
    \end{prooftree}
    
    \begin{prooftree}
    \hypo{T \in dom(\sigma)}
    \hypo{\langle \texttt{mode}(T), t, \sigma \rangle \rightarrow \texttt{read}}
    \hypo{\langle \texttt{mode}(T), t, \sigma' \rangle \rightarrow \texttt{update}(\oplus)}
    \infer3[$Thaw$]{\langle T \texttt{ = thaw}(T, \oplus), t, \sigma \rangle \rightarrow \sigma'}
    \end{prooftree}

    \rulesep

    \begin{prooftree}
    \hypo{\texttt{parent}(D_l) = \texttt{device}(t_g)}
    \hypo{n = \texttt{nthreads}(D_l)}
    \infer2[\hlite{$ParallelFor$}]{\splitfrac{\langle \texttt{for}\ i = \texttt{parallel}(a\texttt{:} b, D_l, h)\ \texttt{do}\ s\ \texttt{end}, t_g, \sigma \rangle \rightarrow }{\langle \texttt{spawn}(\texttt{worker}(i, a\texttt{:} b, h, s), D_l, n), t_g, \sigma[i_{chunk} \mapsto a] \rangle}}
    \end{prooftree}

    \rulesep

    \begin{prooftree}
    \hypo{s' = \texttt{transfer}(T_1, \texttt{transfer}(T_2, ... \texttt{transfer}(T_m, s)...))}
    \hypo{\forall_k, \texttt{fiber}(T_k, ...) \in s}
    \infer2[\hlite{$Spawn$}]{\langle \texttt{spawn}(s, D_l, n), t_g, \sigma \rangle \rightarrow \langle \langle s', \texttt{task}(D_l, 1), \{\} \rangle \| ... \| \langle s', \texttt{task}(D_l, n), \{\} \rangle, t_g, \sigma \rangle}
    \end{prooftree}

    \rulesep

    \begin{prooftree}
    \hypo{\langle \texttt{mode}(T), t, \sigma \rangle \rightarrow \texttt m}
    \hypo{l = \mathop{max}\left(\{k | fiber(T, k, p), \in s\}\right)}
    \infer2[\hlite{$Transfer$}]{\langle \texttt{transfer}(T, s), t, \sigma \rangle \rightarrow \langle \texttt{checkout}(T, l, m, t_l); s; \texttt{commit}(T, l, m, t_l), t_l, \sigma \rangle}
    \end{prooftree}

    \rulesep






    
    \begin{prooftree}
    \hypo{\langle C_1, t, \sigma \rangle \rightarrow \langle C_1', t, \sigma' \rangle}
    \infer1[\hlite{$ParLeft$}]{\langle C_1 \| C_2, t, \sigma \rangle \rightarrow \langle C_1' \| C_2, t, \sigma' \rangle}
    \end{prooftree}
    \hfill
    \begin{prooftree}
    \hypo{\langle C_2, t, \sigma \rangle \rightarrow \langle C_2', t, \sigma' \rangle}
    \infer1[\hlite{$ParRight$}]{\langle C_1 \| C_2, t, \sigma \rangle \rightarrow \langle C_1 \| C_2', t, \sigma' \rangle}
    \end{prooftree}

    \rulesep

    \begin{prooftree}
    \hypo{\langle y, t_l, \sigma_l \rangle \rightarrow v}
    \infer1[\hlite{$WriteLocal$}]{\langle x=y, t_l, \sigma_l \rangle \rightarrow \langle \texttt{skip}, t_l, \sigma_l[x \mapsto v] \rangle}
    \end{prooftree}

    \rulesep

    \begin{prooftree}
    \hypo{\langle y, t_l, \sigma_l \rangle \rightarrow v}
    \hypo{x \not \in dom(\sigma_l)}
    \infer2[\hlite{$WriteGlobal$}]{\langle \langle x=y, t_l, \sigma_l \rangle, t_g, \sigma_g \rangle \rightarrow \langle x=v; \langle \texttt{skip}, t_l, \sigma_l \rangle, t_g, \sigma_g \rangle}
    \end{prooftree}

    \rulesep

    \begin{prooftree}
    \hypo{\sigma_l[x] = v}
    \infer1[\hlite{$ReadLocal$}]{\langle x, t_l, \sigma_l \rangle \rightarrow v}
    \end{prooftree}
    \hfill
    \begin{prooftree}
    \hypo{\langle x, t_g, \sigma_g \rangle \rightarrow v}
    \hypo{x \not \in dom(\sigma_l)}
    \infer2[\hlite{$ReadGlobal$}]{\langle \langle x, t_l, \sigma_l \rangle, t_g, \sigma_g \rangle \rightarrow \langle v, t_l, \sigma_l \rangle}
    \end{prooftree}

    \rulesep

    \begin{prooftree}
    \hypo{\texttt{nthreads}(\texttt{device}(t)) = n}
    \hypo{\texttt{threadid}(t) = q}
    \hypo{\texttt{kind}(h) = \texttt{static}}
    \infer3[\hlite{$StaticFor$}]{\langle \texttt{worker}(i, a\texttt{:} b, h, s), t, \sigma \rangle \rightarrow \left\langle \splitfrac{\texttt{for}\ i = a + \lfloor (q-1)(b-a+1)/n \rfloor \texttt{:}}{a + \lfloor q(b-a+1)/n \rfloor - 1\ \texttt{do}\ s\ \texttt{end}, t, \sigma} \right\rangle}
    \end{prooftree}

    \rulesep

    \begin{prooftree}
    \hypo{j = \sigma_g[i_{chunk}]}
    \hypo{j \leq b}
    \hypo{\texttt{kind}(h) = \texttt{dynamic}}
    \hypo{\texttt{step}(h) = c}
    \infer4[\hlite{$DynamicFor$}]{\splitfrac{\langle\langle \texttt{worker}(i, a\texttt{:} b, h, s), t_l, \sigma_l \rangle, t_g, \sigma_g\rangle \rightarrow}{\left \langle \left \langle \splitfrac{\texttt{for}\ i = j\texttt{:} \min(j+c-1, b)\ \texttt{do}\ s\ \texttt{end};}{\texttt{worker}(i, a\texttt{:} b, h, s), t_l, \sigma_l} \right \rangle, t_g, \sigma_g[i_{chunk} \mapsto j+c]\right \rangle }}
    \end{prooftree}

    \caption{A small-step semantics for WingSpan's parallel and serial for nodes. Variable domains are described in Figure \ref{fig:syntax}. Context triples are written as $C = \langle s, t, \sigma\rangle$, where $s$ is the program to execute, $t$ is the task on which it runs, and $\sigma$ is the environment. We nest context triples inside the program to represent local scopes or child tasks. We use $s[x\mapsto y]$ to express state modifications and program rewrites.
    The $SerialFor$ rule is a stand-in for structured coiteration \cite{ahrens_looplets_2023}. Highlighted rules are new concepts added in this paper.}

    \label{fig:semantics}
\end{figure}

\subsection{Building on Finch}

Our semantics build on those of Finch \cite{ahrens_finch_2025}, and apply only to concordant loops (where the subscripts on all accesses are bare loop indices in the same order as the loops are nested). Note that it is always possible to convert a non-concordant program to a concordant one by inserting single-iteration loops. The rules from $Call$ to $Thaw$ represent a simplified semantics for how Finch lowers structured loops one-by-one, descending through levels as corresponding loop indices are reached. The $Unfurl$ rule introduces a \texttt{looplet} dimension, which is a shorthand for structured coiteration using looplets \cite{ahrens_looplets_2023} and is omitted for brevity. In situations where threads iterate a subset of the full range, looplets define methods to efficiently seek through each level to the start of the range.

Tensors in Finch flow through a few lifecycle functions for initialization and finalization. Each level may be in one of three modes: \texttt{undeclared}, \texttt{read}, and $\texttt{update}(\oplus)$, where $\oplus$ is the reduction operator in use. Levels may only change modes in the scope in which they are declared. Three functions change modes; \texttt{declare} initializes to \texttt{update} mode, \texttt{freeze} finalizes to \texttt{read} mode, and \texttt{thaw} switches to \texttt{update} mode without initializing to the fill value.

\subsection{Parallel Loops, Devices and Transfers}

WingSpan introduces several strategies to avoid data races, from locking fiber subtrees to replicating levels across different processors. Some of these features require the data structure to make device-specific assumptions. For example, the merge level makes a copy for each thread.
Additionally, loops may be parallelized with different thread counts, and nested parallelism requires different guarantees than flat parallelism. Each loop must be associated with a device. To coordinate these concerns, we introduce a lightweight abstraction for \textbf{devices} and the \textbf{threads} that run on them.

In our device hierarchy, each thread has a \textbf{\texttt{threadid}}, a \textbf{\texttt{device}}, and \textbf{\texttt{parent}} thread which spawned it. Each device has a unique symbolic identity, a \textbf{\texttt{parent}} device, and a \textbf{\texttt{num\hyp threads}} it spawns. To specify a parallel loop, we wrap the iteration domain in a \textbf{\texttt{parallel}} construct ($Parallel$ rule).
When we spawn a new thread, we \textbf{\texttt{transfer}} all of the tensors used in the loop body to the thread's device ($Spawn$ rule). The \textbf{\texttt{checkout}} and \textbf{\texttt{commit}} functions are interfaces that each level must implement in order to transfer its underlying data buffers to the new parallel region. Each function receives the tensor, the new device to transfer to, and the maximum level currently used by the tensor ($Transfer$ rule). Note that the entire suffix of levels is transferred each time. Some levels are \textbf{device-gated}, meaning that they can only be checked out or unfurled on a particular device. The transfer functions have different roles depending on the tensor mode. If the tensor is \texttt{undeclared}, it is local to the parallel region, so each thread is assigned a copy. If the tensor is in \texttt{read} mode, we only need to make a device-readable handle for it. When the tensor is in \texttt{update} mode the transfer is more complex because we need to coordinate a shared output.

\subsection{Modifier Levels for Concurrency}

The combination of different level formats supported by Finch, different concurrency situations, and different resolution strategies represents a significant implementation burden. To orthogonalize concerns, we implement several \textbf{modifier} levels which add safety guarantees to existing formats by replicating or locking subfibers. The \textbf{mutex} level spinlocks each subfiber, the \textbf{isolate} level allocates each subfiber separately, the \textbf{shard} level makes a separate level for each thread, and the \textbf{merge} level creates copies and recombines them at the end of the parallel updates.

\begin{figure}[H]
    \centering
    \includegraphics[width=0.95\linewidth]{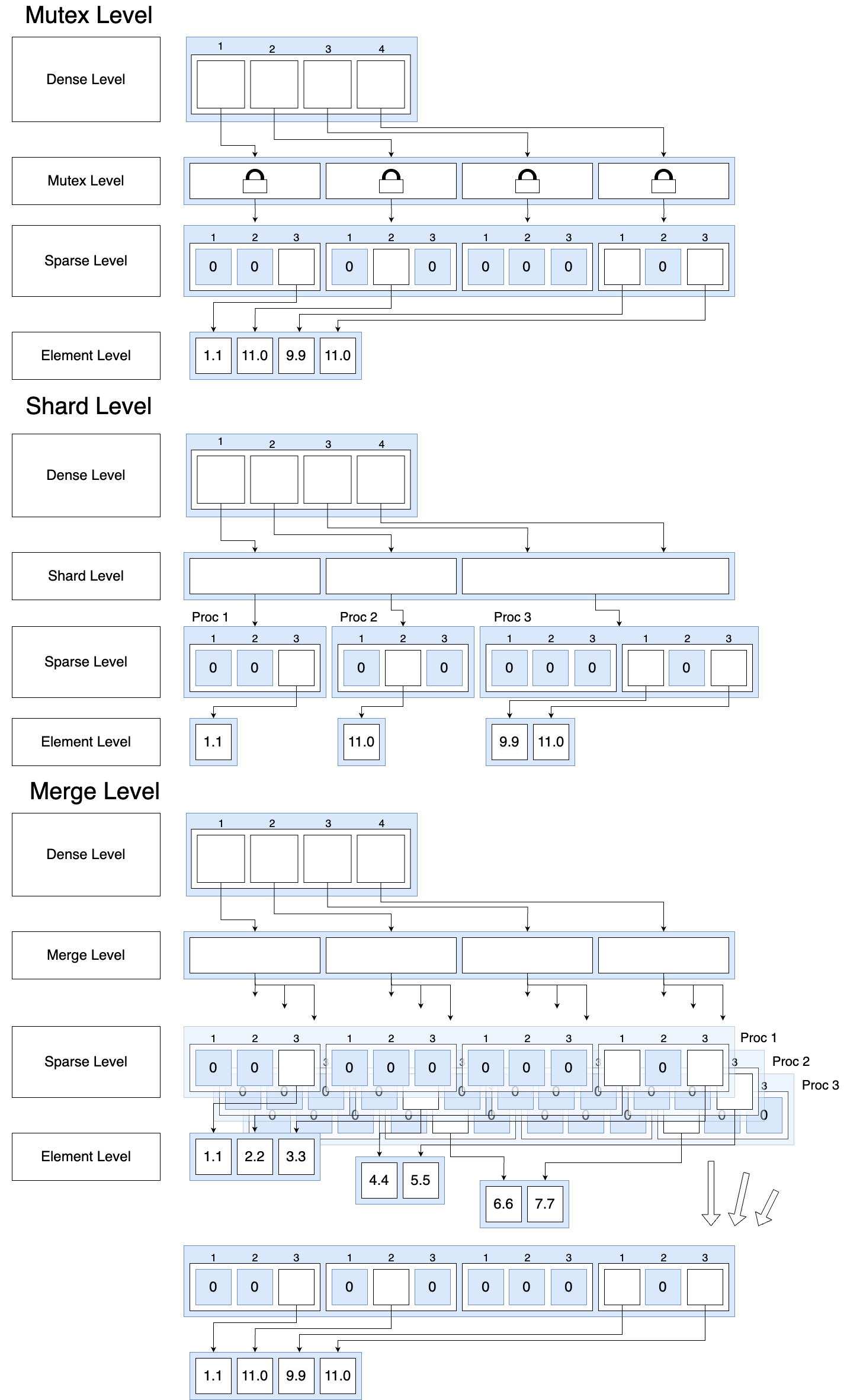}
    \Description{Three level formats proposed by WingSpan. The Mutex Level places a lock on all of the positions in its child level. The Shard Level uniquely assigns each child position to one and only one processor. Finally, the Merge Level clones the child level and gives each processor a copy. The processors write to their copies in isolation and ultimately recombine them using a parallel merging algorithm at the end of computation.}
    \caption{Three proposed level formats to handle concurrent updates. When processors write to separate locations, we can split the data structure across processors with the shard level. When processors write to the same locations, we need to use locks or keep different copies on each processor and merge them at the end of computation.}\label{fig:parallel-levels}
\end{figure}

Unlike traditional levels, modifier levels do not introduce new modes into the tensor. Instead, they wrap the behavior of existing level access functions like \texttt{unwrap}, \texttt{increment}, and \texttt{unfurl}, which get called just before we iterate over or access a level (rules $Access$, $Assign$, and $Unfurl$). Because modifier levels make device-specific assumptions, their stability guarantees are also device-specific. We introduce the notation $S^D_k$ to describe the stability of a level with respect to threads on device $D$. This notation comes at no loss of generality to our dependence test, since a parallel loop's device can easily be supplied in addition to its index. The stability of each modifier is given in Figure \ref{fig:stability}.

\begin{figure}
\begin{align*}
S_{l+1} &= \{node, sibling, cousin\} \quad \quad &(Dense)\\
S_1 &= \{sibling, cousin\} \quad \quad &(Element)\\
S_1 &= \{node, sibling, cousin\} \quad \quad &(Atomic)\\
S_{l+1} &= \varnothing \quad \quad &(Sparse, etc.)\\
S'^D_k &= S^D_k \cup \{node, sibling\} \quad k \leq l+1 \quad &(Mutex) \\
S'^D_k &= S^D_k \cup \{cousin\} \quad k \leq l+1 \quad &(Isolate) \\
S'^D_k &= S^D_k \cup \{cousin\} \quad k \leq l+1 \quad &(Shard(D)) \\
S'^D_k &= S^D_k \cup \{node, sibling, cousin\} \quad k \leq l+1 \quad &(Merge(D))
\end{align*}
\caption{The stability $S'$ of each level at mode $l$, expressed in terms of the stability $S$ of the wrapped levels when applicable. Only devices which \texttt{unwrap}, \texttt{increment}, or \texttt{unfurl} the modifier may use the modifier for stability.}
\label{fig:stability}
\end{figure}


The \textbf{mutex level} creates a lock on every fiber in its child. The lock is taken on invocation of \texttt{unwrap}, \texttt{increment}, or \texttt{unfurl}. The level itself is simply a list of these locks, any of which can be accessed in parallel by any combination of threads or devices, but stability is only provided on devices that take the locks. Instead of using \texttt{Mutex(Element(0.0))}, we also offer an \textbf{atomic} level which is a leaf level guarded by system atomics, which are faster than spin-locks for scalars.

The \textbf{isolate level} allocates a separate sublevel for each child. This allows different threads to modify different subtrees concurrently. Like the mutex level, the isolate level cannot protect against parallelism from devices which are spawned after the isolated subtree is loaded. The combination of \texttt{Mutex(Isolate(...))} can guard against any race conditions from containing loops.

The \textbf{shard level} improves on the isolate level significantly by allocating separate sublevels ("shards") only once per thread, rather than once per child. Shard levels are gated by a device, denoted \textit{Shard(D)}, that defines the number of shards to initialize. Shard levels impose the additional constraint that different positions will be written to by different threads. The fibers in the child level are then claimed by a single, unique thread on the first write. In Figure \ref{fig:parallel-levels}'s sharded structure, the shard level's device contains three threads. The first thread has claimed fiber 1 in the sparse level, the second thread fiber 2, and the third thread the remaining two fibers. Each thread stores its claimed fibers as an isolated subtree of the overall tensor tree, enabling a parallel algorithm to access distinct fibers at the same time.

The \textbf{merge level} enables parallelism with fewer restrictions by cloning the entire sublevel for each thread on its device. Merge levels are device-gated, and since each thread operates on its own copy, race conditions are eliminated. During the \texttt{freeze} step, a per-level \textbf{\texttt{coalesce}} routine recombines the disparate copies into a unified result. This routine is self-contained---it requires no knowledge of parent or child formats---and runs in time proportional to the total number of non-zeroes, with non-zeroes partitioned across available threads for load balance. The interface is $coalesce(local, global, task, lvl, P)$, where $P$ is the number of processors, $lvl$ is the level to merge, $local$ is the list of $lvl$'s positions on each processor, $global$ is the merged positions, and $task$ maps each fiber $local[i]$ on processor $task[i]$ to fiber $global[i]$ in the output. A level computes its child's metadata during \texttt{coalesce}, allowing the procedure to compose across the tensor tree.

\begin{figure*}
    \centering

    \begin{subfigure}{0.5\linewidth}
        \begin{minted}[fontsize=\scriptsize,breaklines,escapeinside=||]{python}
|\colorbox{serial_blue}{dev1 = cpu(:d1, 4)}|
|\colorbox{serial_blue}{dev2 = cpu(:d2, 4)}|
|\colorbox{serial_blue}{C = Tensor(Dense(Shard(dev1, Dense(Shard(dev2, Element(0.0))))))}|
|\colorbox{serial_blue}{A = Tensor(Dense(Dense(Element(0.0))))}|
|\colorbox{serial_blue}{B = Tensor(Dense(Dense(Element(0.0))))}|

|\colorbox{serial_blue}{C .= 0}|
|\colorbox{serial_blue}{\texttt{for j=parallel(\_, dev1, static\_schedule())}}|
    |\colorbox{dev1green}{\texttt{for i=parallel(\_, dev2, static\_schedule())}}|
        |\colorbox{dev2yellow}{\texttt{for k=\_}}|
            |\colorbox{dev2yellow}{C[i, j] += A[i, k] * B[k, j]}|
        |\colorbox{dev2yellow}{end}|
    |\colorbox{dev1green}{end}|
|\colorbox{serial_blue}{end}|
        \end{minted}
    \end{subfigure}%
    \hfill
    \begin{subfigure}{0.5\linewidth}
        \centering
        \includegraphics[width=\linewidth]{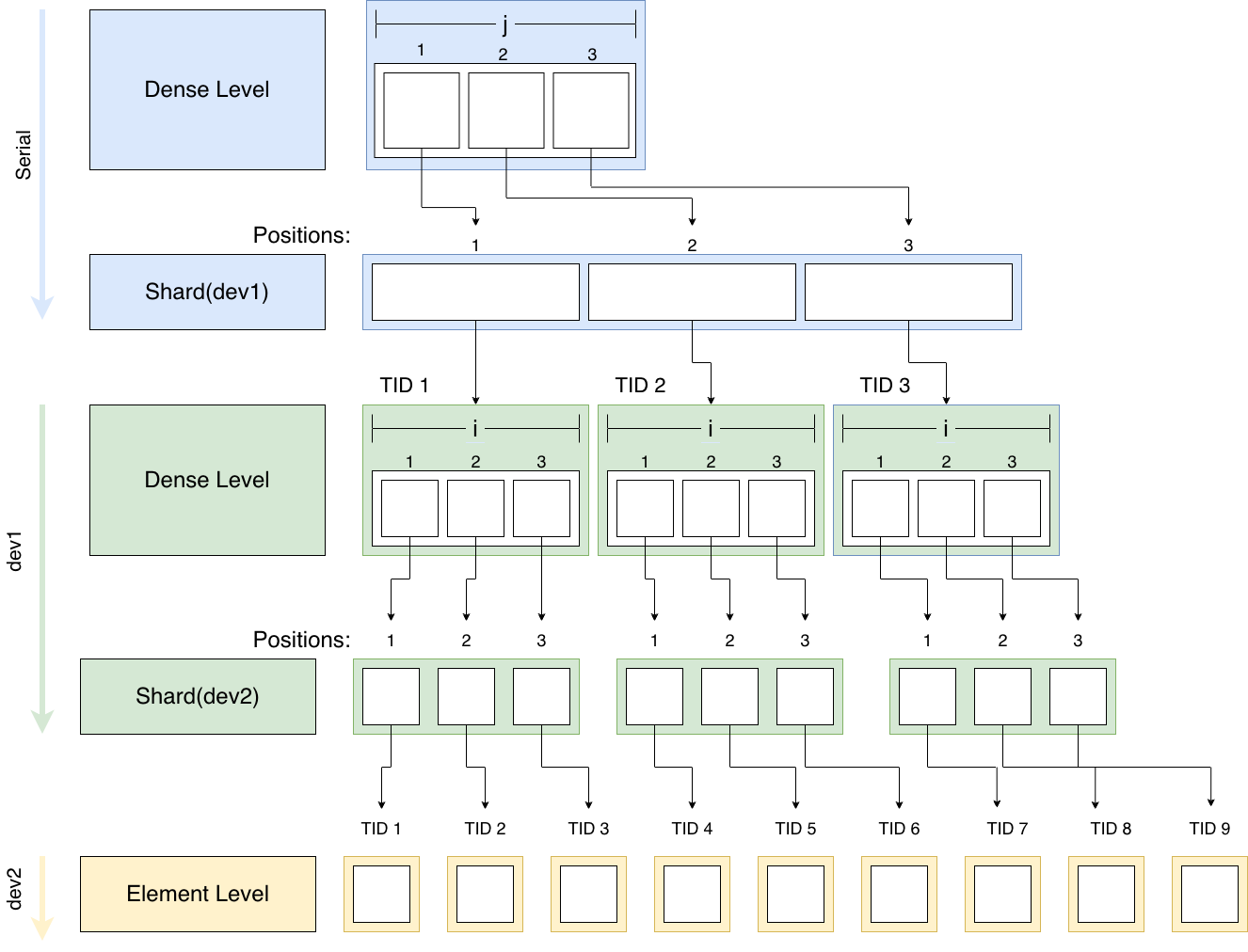}
    \end{subfigure}

    \caption{An annotated nested inner product (left) and its output $C$ (right). Blue, green, and yellow highlight code running on the serial device, \texttt{dev1}, and \texttt{dev2}, respectively. Entering a parallel loop transfers in-scope tensors to that loop's device. The tree shows $C$'s structure on each device: a vector of vectors (serial), a vector of elements (\texttt{dev1}), or a single element (\texttt{dev2}).}
    \label{fig:nested_para}
\end{figure*}

Because devices can be nested, device-gated levels may enforce certain loop orderings. In Figure \ref{fig:nested_para}, the loop over \texttt{dev2} must be nested within \texttt{dev1}'s loop since it expects to transfer from a vector of scalars to a single scalar. Reversing this order causes \texttt{dev2} to encounter an unexpected vector of vectors, yielding an undefined state. Nested device-gated levels must be parallelized in the order they appear in the tensor tree. As a corollary, device-gated levels must be unfurled by the parallel loop over their device. This distinguishes them from non-gated levels, which can be accessed by any parallel loop.

\subsection{Schedules and Load Balancing}

WingSpan introduces two thread scheduling algorithms for its parallel nodes. The \texttt{static} schedule evenly divides the iteration space of the loop by the number of active threads (Rule $Static$). This node schedule is simple to implement, but can be slow when work is not evenly balanced across the partitions. To address load balancing concerns, the \texttt{dynamic} algorithm splits the iteration space using a user-defined chunk size (Rule $Dynamic$). Each thread claims chunks dynamically until the iteration space is exhausted, allowing quickly-finishing threads to acquire new work instead of stalling. Both schedulers allocate local tensors and perform transfers once per thread startup and shutdown, not once per loop iteration (which would be the default behavior unless tensor-aware schedulers were implemented).

\section{WingSpan Performance}

Our results make three claims distinguishing WingSpan from similar software. First, WingSpan is general, expressing a variety of parallel patterns on arbitrary tensor kernels where comparable software may lack parallelism or an efficient routine. Second, this generality does not sacrifice performance; WingSpan competes with specialized, hand-written routines. Finally, WingSpan handles structure beyond sparsity, supporting parallelism over tensors with run-length patterns.

For reproducibility, we use a 16-core subset of dual Intel Xeon 6972P processors, running at 2.4 GHz with a 480 MB L3 cache. All runtime measurements are the minimum of 10,000 warm-cache trials or 5 seconds of measurement, whichever happens first. We use a combination of Red Hat Enterprise Linux 9, Eigen version 5.0.0, MKL version 2025.3.1.11, SuiteSparseGraphBLAS v0.11.0, and TACO commit 1278503 to evaluate \cite{guennebaud_eigen_2010, noauthor_developer_2024, 10.1145/3322125, kjolstad_taco_2017}. Read-only inputs have type \texttt{Dense(\allowbreak SparseList(\allowbreak Element(0.0)))} unless explicitly stated otherwise. We evaluate using dynamic and static load balancers and present the best schedule.

\subsection{Sparse Matrix-Sparse Matrix Multiply (SpGEMM)}

This benchmark computes the matrix product $C=A\cdot B$ where $A,B,$ and $C$ are all sparse. We compare WingSpan's kernels to equivalent MKL, GraphBLAS, and Eigen code.

\begin{figure}
    \begin{minipage}{0.5\linewidth}
        \includegraphics[width=\columnwidth]{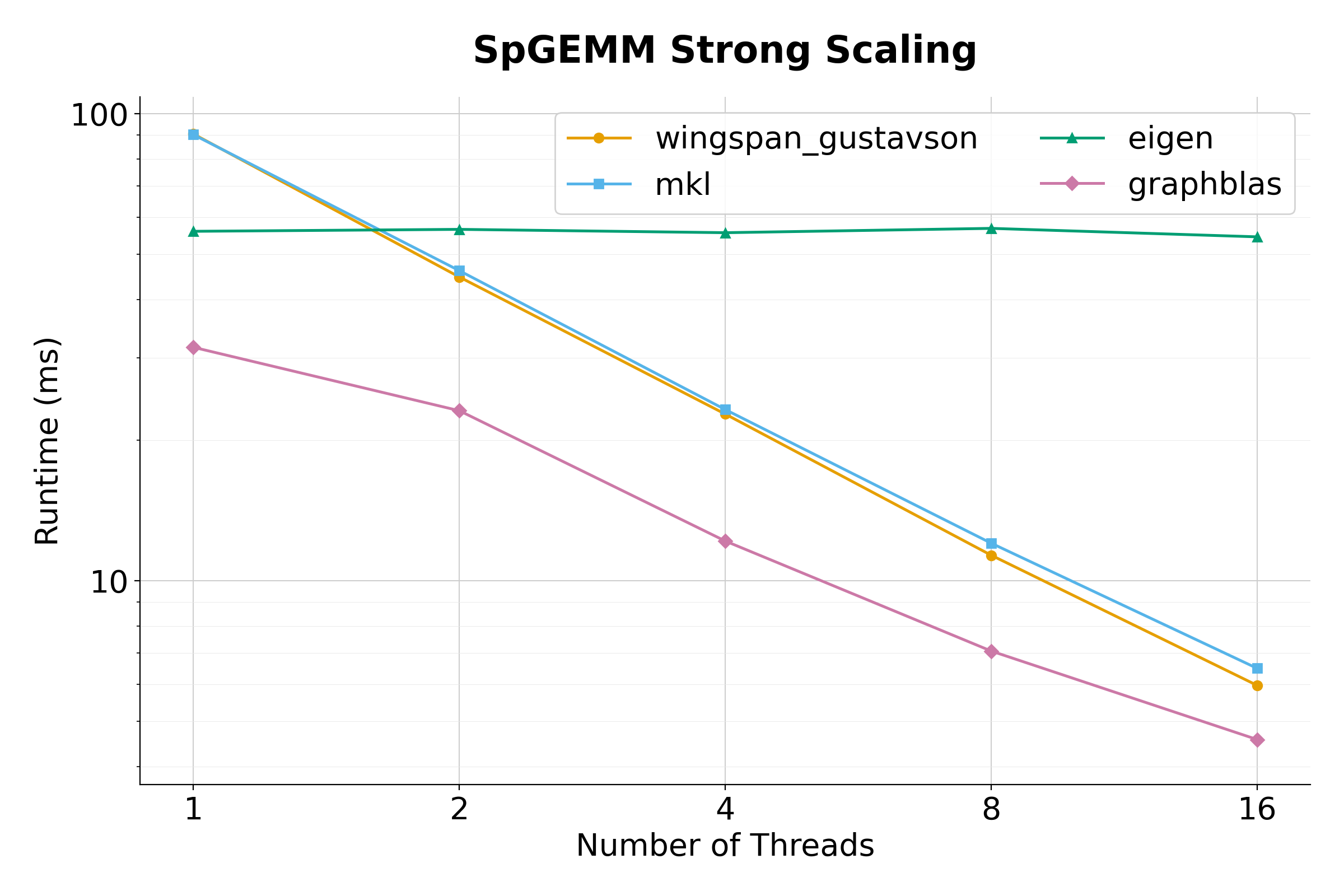}
    \end{minipage}%
    \begin{minipage}{0.5\linewidth}
        \includegraphics[width=\columnwidth]{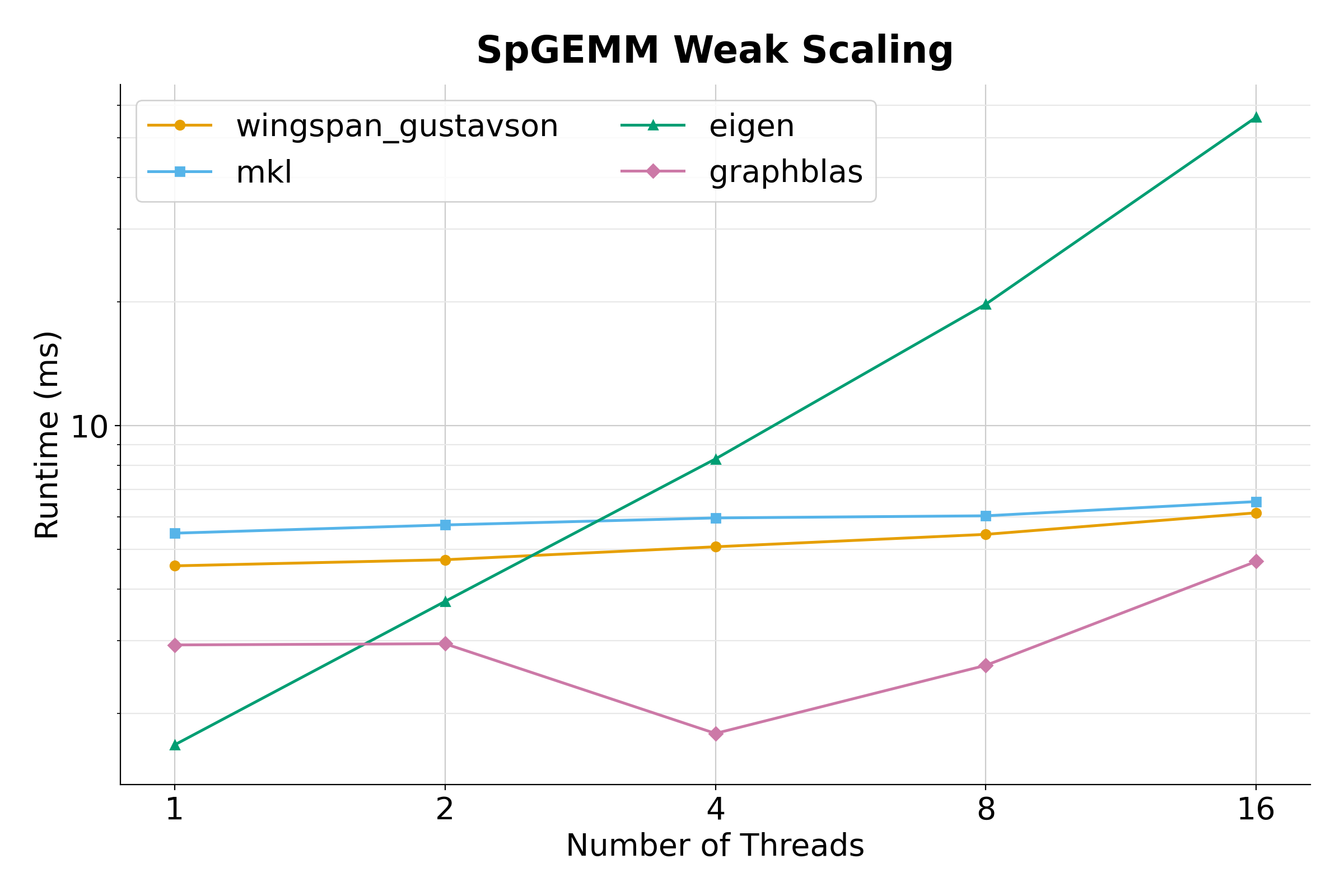}
    \end{minipage}
    \caption{SpGEMM strong and weak scaling on randomly generated sparse matrices with 12 non-zeroes per row.}
    \label{fig:spgemm_scale}
\end{figure}

\begin{figure*}[t]
    \centering
    \begin{subfigure}{\columnwidth}
        \includegraphics[width=\columnwidth]{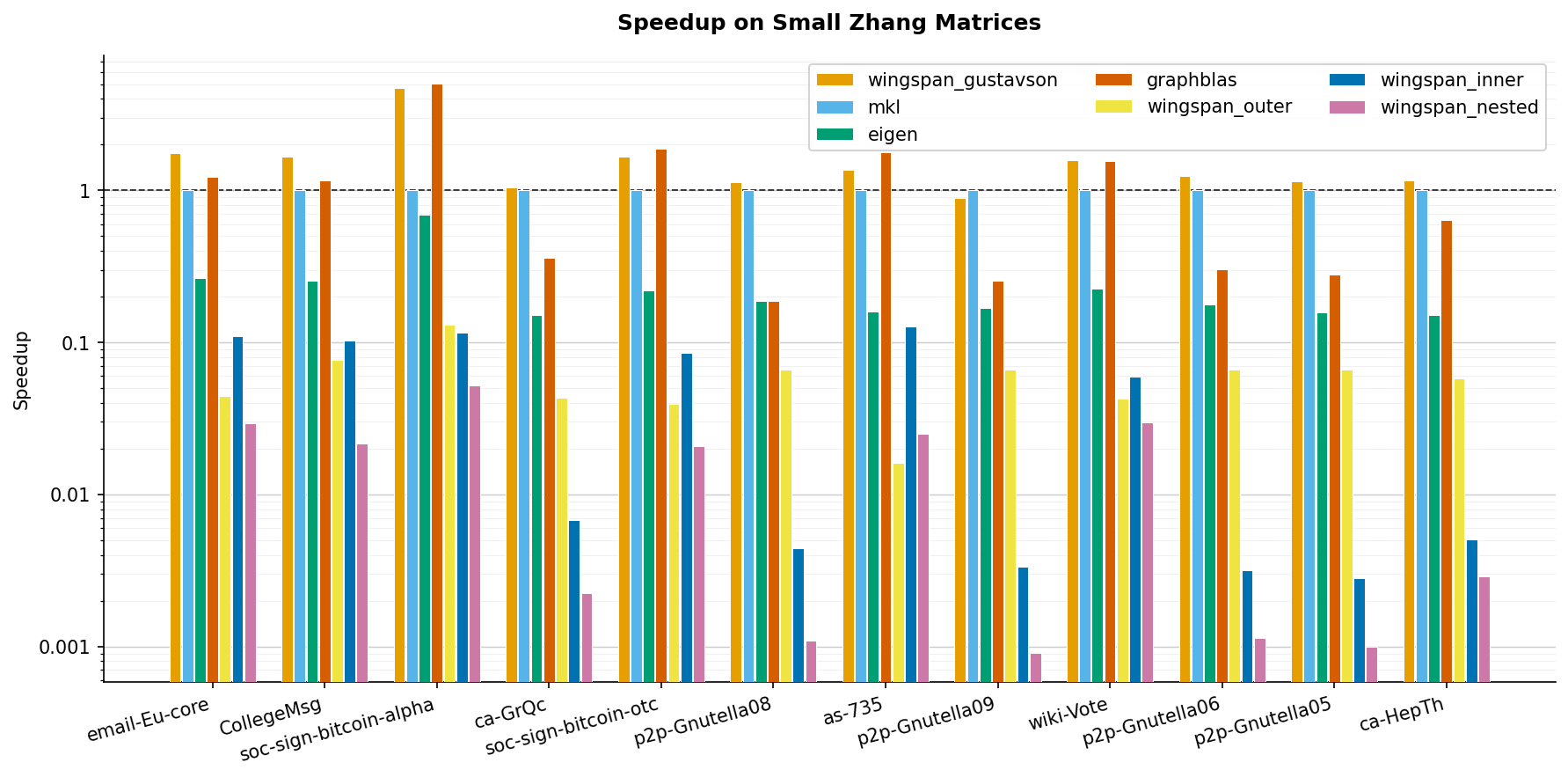}
    \end{subfigure}
    \begin{subfigure}{\columnwidth}
        \includegraphics[width=\columnwidth]{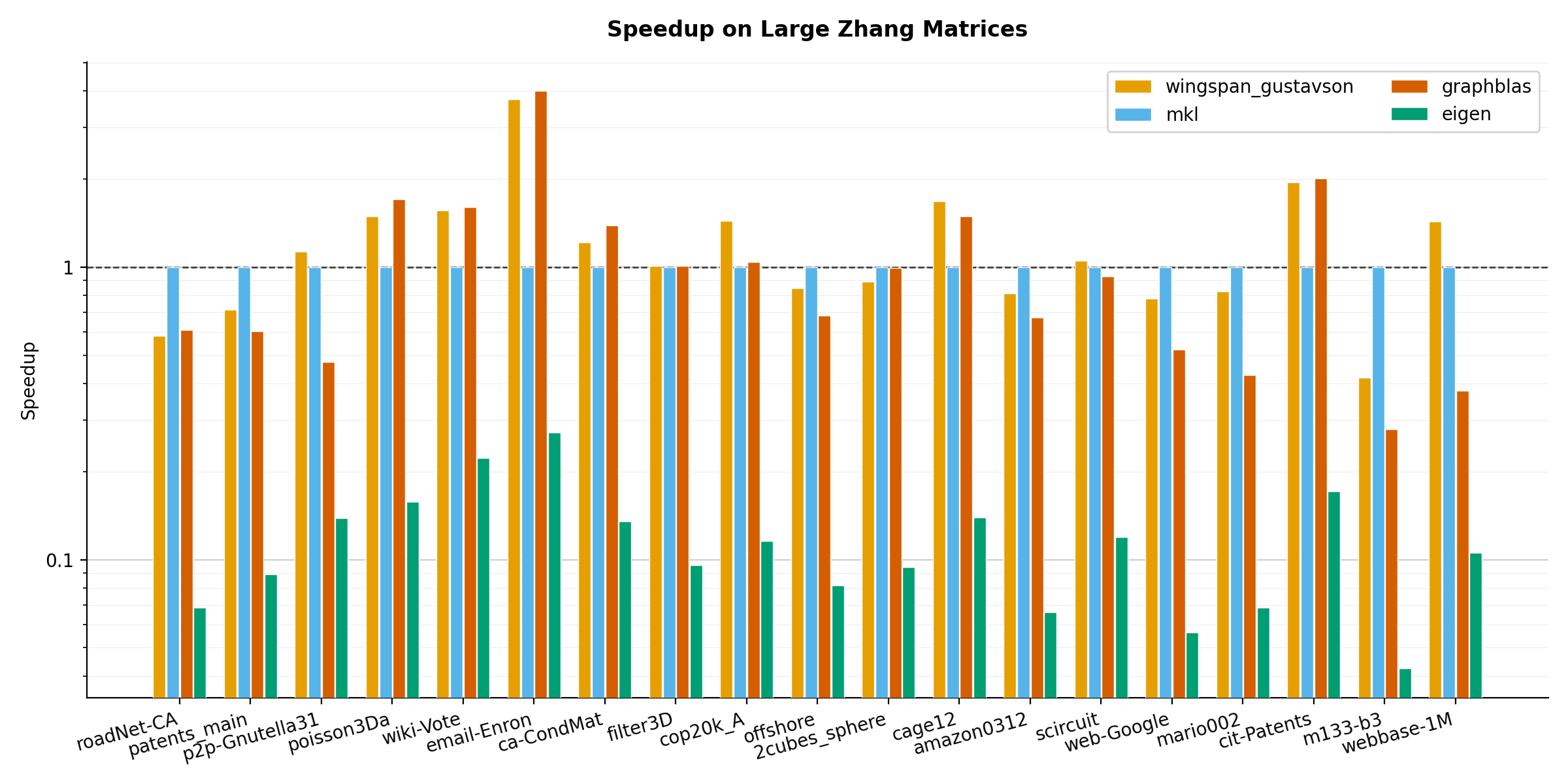}
    \end{subfigure}

    \caption{SpGEMM speedup on the small (left) and large (right) Zhang matrices.}
    \label{fig:spgemm_zhang}
\end{figure*}

\begin{figure}[t]
\centering
\scriptsize
\begin{minipage}[t]{0.5\linewidth}
\textbf{Inner Product}
\begin{minted}[]{julia}
dev=cpu(:t,nt)
C=Tensor(Dense(Shard(
  dev,SparseList(Element(0.0)))))
C.=0
for j=parallel(_,dev,sch)
  for i=_
    for k=_
      C[i,j]+=AT[k,i]*B[k,j]
\end{minted}
\end{minipage}%
\hfill
\begin{minipage}[t]{0.5\linewidth}
\textbf{Outer Product}
\begin{minted}[]{julia}
dev=cpu(:t,nt)
C=Tensor(Merge(dev,SparseDict(
  SparseDict(Element(0.0)))))
C.=0
for k=parallel(_,dev,sch)
  for j=_
    for i=_
      C[i,j]+=A[i,k]*BT[j,k]
\end{minted}
\end{minipage}%

\vspace{11pt}

\begin{minipage}[t]{0.5\linewidth}
\textbf{Gustavson's Algorithm}
\begin{minted}[]{julia}
dev=cpu(:t,nt)
C=Tensor(Dense(Shard(dev,
  SparseList(Element(0.0)))))
w=Tensor(SparseByteMap(Element(0.0)))
C.=0
for j=parallel(_,dev,sch)
  w.=0
  for k=_
    for i=_
      w[i]+=A[i,k]*B[k,j]
  for i=_
    C[i,j] =w[i]
\end{minted}
\end{minipage}%
\hfill
\begin{minipage}[t]{0.5\linewidth}
\textbf{Nested Parallelism (Gustavson's)}
\begin{minted}[]{julia}
dev1=cpu(:t,isqrt(nt))
dev2=cpu(:q,isqrt(nt))
C=Tensor(Dense(Shard(dev1,
  SparseList(Element(0.0)))))
w=Tensor(Merge(dev2,
  SparseByteMap(Element(0.0))))
C.=0
for j=parallel(_,dev,sch)
  w.=0
  for k=parallel(_,dev2)
    for i=_
      w[i]+=A[i,k]*B[k,j]
  for i=_
    C[i,j]=w[i]
return C
\end{minted}
\end{minipage}
\caption{SpGEMM implemented in WingSpan. For brevity, trailing \texttt{end} scope-closing annotations are omitted.}
\label{fig:spgemm_allkernels}
\end{figure}

To demonstrate WingSpan's flexibility, we implement four SpGEMM algorithms in the language, shown in Figure \ref{fig:spgemm_allkernels}. Our dependence theory informs the choice of parallel levels made in each kernel. The inner products, Gustavson's, and nested kernels all parallelize the $j$ loop, which means the access to C$[i,j]$ will always have unique positions. Therefore, only cousin dependencies occur and we use the shard level. In the outer product, the parallelization of $k$ will cause threads to access both identical and distinct C$[i,j]$. We use a merge level to protect against all three forms of dependence. In the same vein, the workspace in the nested kernel uses a merge level to guard itself from the parallel $k$ accesses.

We test on the matrices used by Zhang et al. in their study of SpGEMM \cite{zhang_gamma_2021}. To optimize for performance, we compare only Gustavson's algorithm on the large matrices. We additionally compute strong and weak scaling plots to verify WingSpan has low parallelization overhead. Figure \ref{fig:spgemm_scale} displays the scaling and Figure \ref{fig:spgemm_zhang} the speedup results.

These results demonstrate WingSpan's flexibility and performance. Figure \ref{fig:spgemm_allkernels}'s wide range of data structures and access patterns show support for nested sparse outputs, scattering write patterns, thread-local parallel accumulators, and nested parallelism. These features can be arbitrarily composed within each other; for example, consider the parallel thread-local accumulator \texttt{w} in the nested kernel. WingSpan's versatility does not trade off with performance. Our Gustavson's kernel attains a geomean $1.21\times$ speedup over MKL and a $1.50\times$ speedup over GraphBLAS across the Zhang matrices. WingSpan's linear strong scaling and near constant weak scaling show it parallelizes with little overhead.

MKL outperforms WingSpan on the outlier roadnet and m133 matrices. These products are large and particularly sparse, even among the other SpGEMM outputs. Because MKL is closed-source, we can only hypothesize that MKL may switch to either a hash accumulator or an expand-sort-contract strategy when the output has a sufficiently low nonzero count, which could improve performance.

\subsection{Sparse Linear Algebra}

We evaluate WingSpan on a variety of linear algebra kernels to assess the applicability of its framework. We consider elementwise addition (SpAdd) and the Hadamard product. We evaluate on the large SNAP matrices from Zhang et al. To benchmark a variety of sparsity patterns, we test both an identity case, where $A=B$, and a standard case, where $B$ is a randomly shuffled copy of $A$. In both cases, we can partition distinct columns to distinct threads, enabling unique positions when accessing C. For the same logic as the inner product SpGEMM kernel, we use the sharded sparse list strategy for both SpAdd and the Hadamard product.

\begin{figure*}[t!]
    \centering
    \begin{subfigure}{\columnwidth}
        \includegraphics[width=\columnwidth]{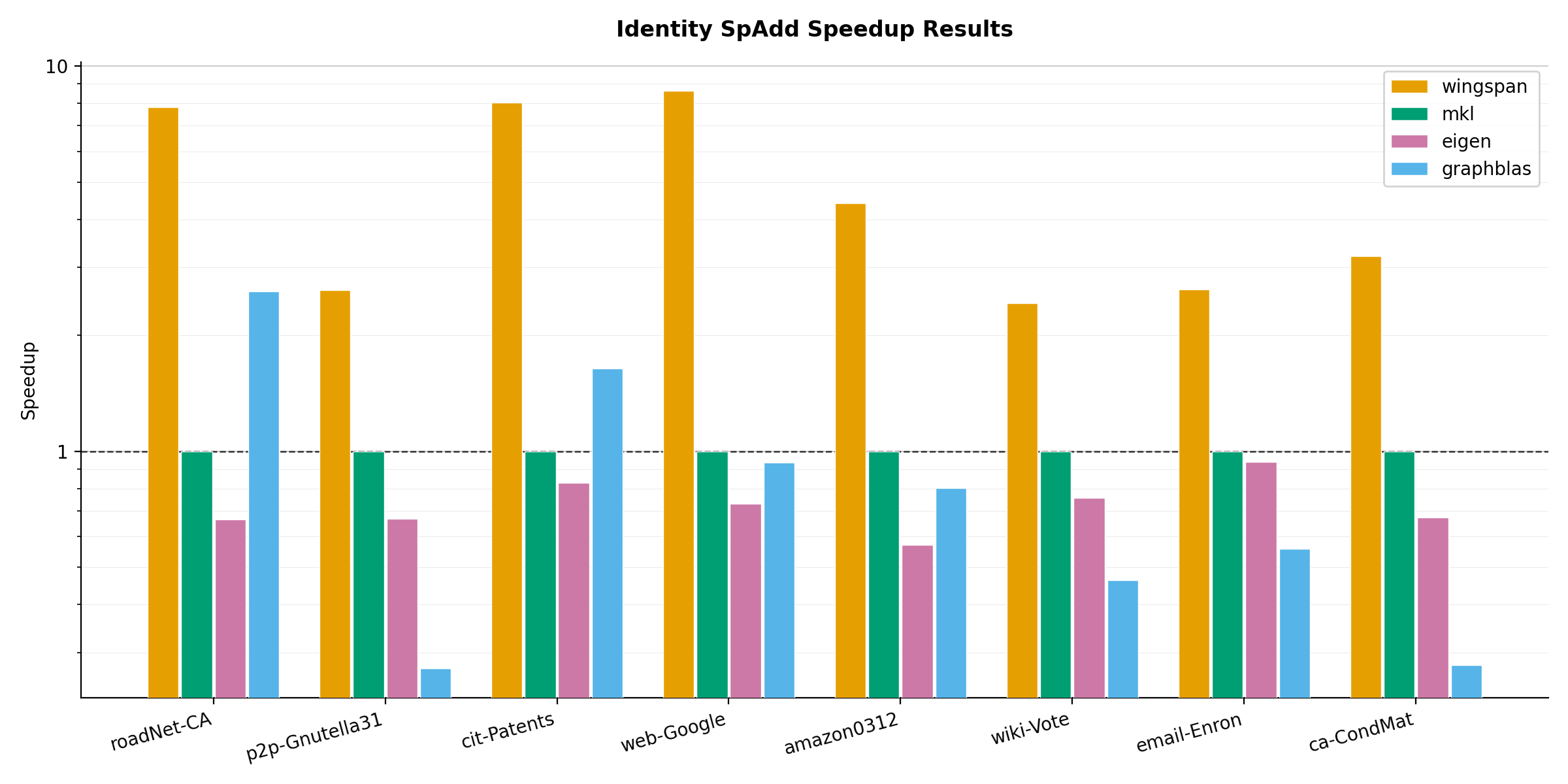}
    \end{subfigure}
    \begin{subfigure}{\columnwidth}
        \includegraphics[width=\columnwidth]{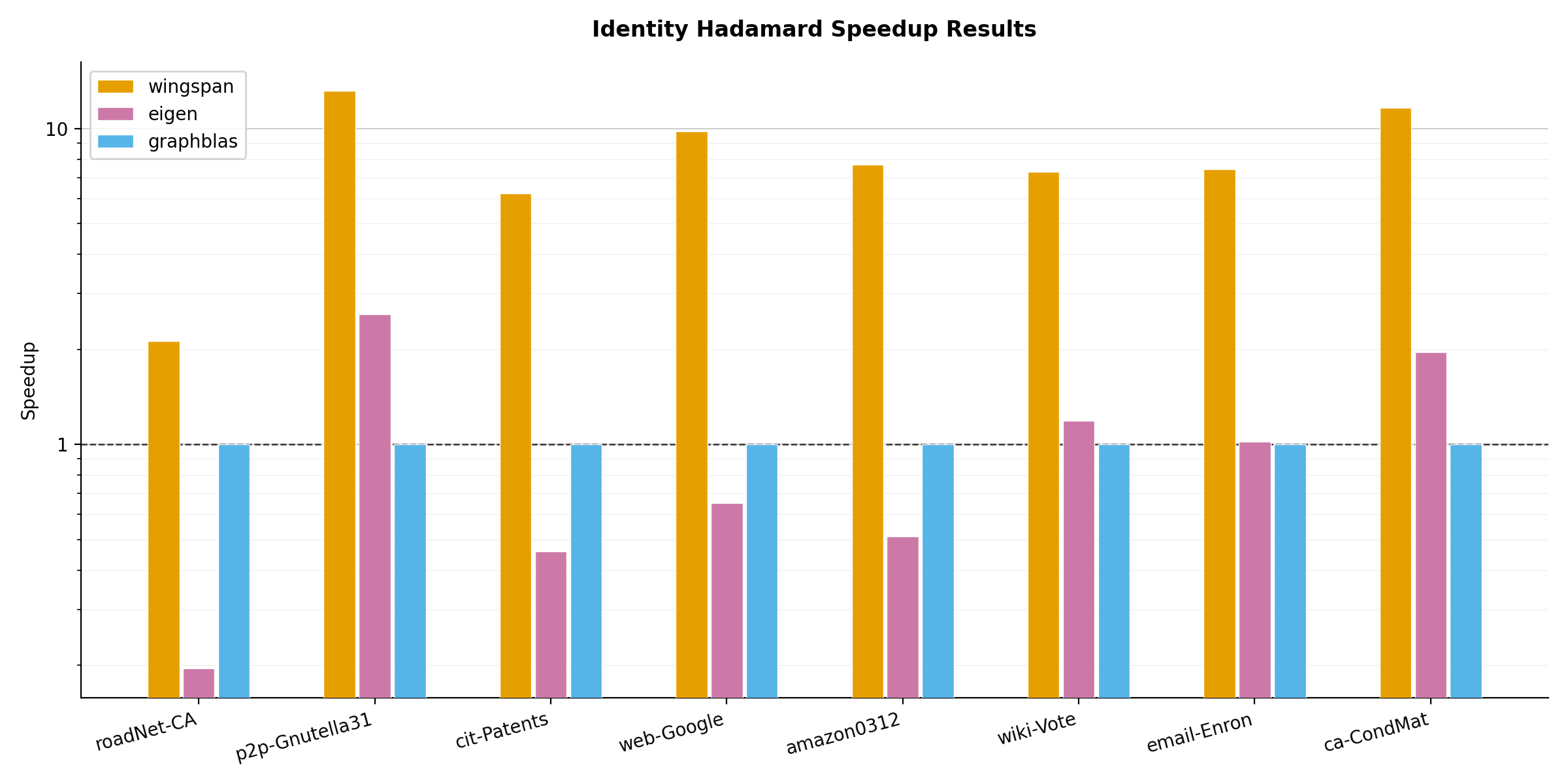}
    \end{subfigure}
    \begin{subfigure}{\columnwidth}
        \includegraphics[width=\columnwidth]{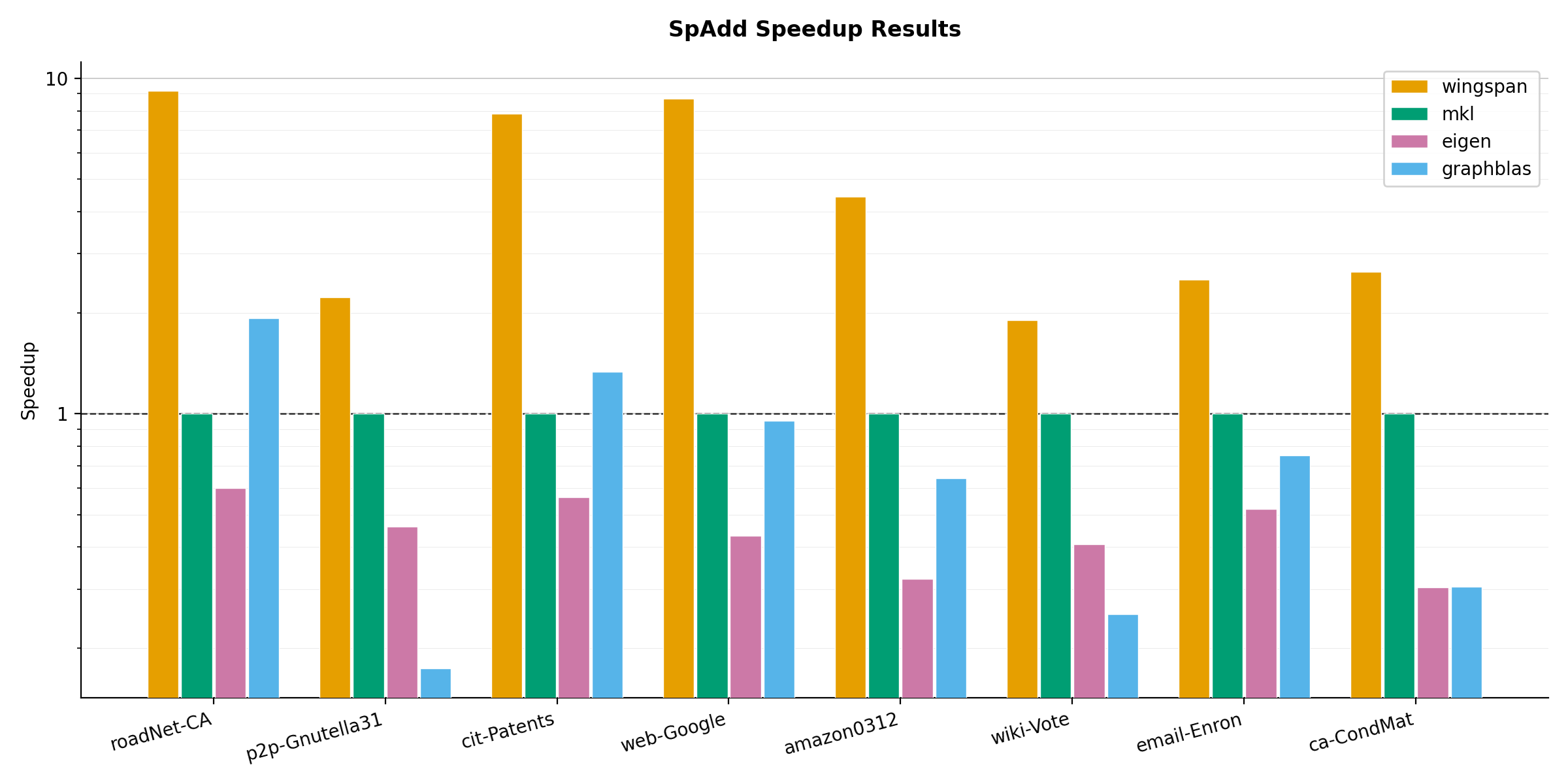}
    \end{subfigure}
      \begin{subfigure}{\columnwidth}
        \includegraphics[width=\columnwidth]{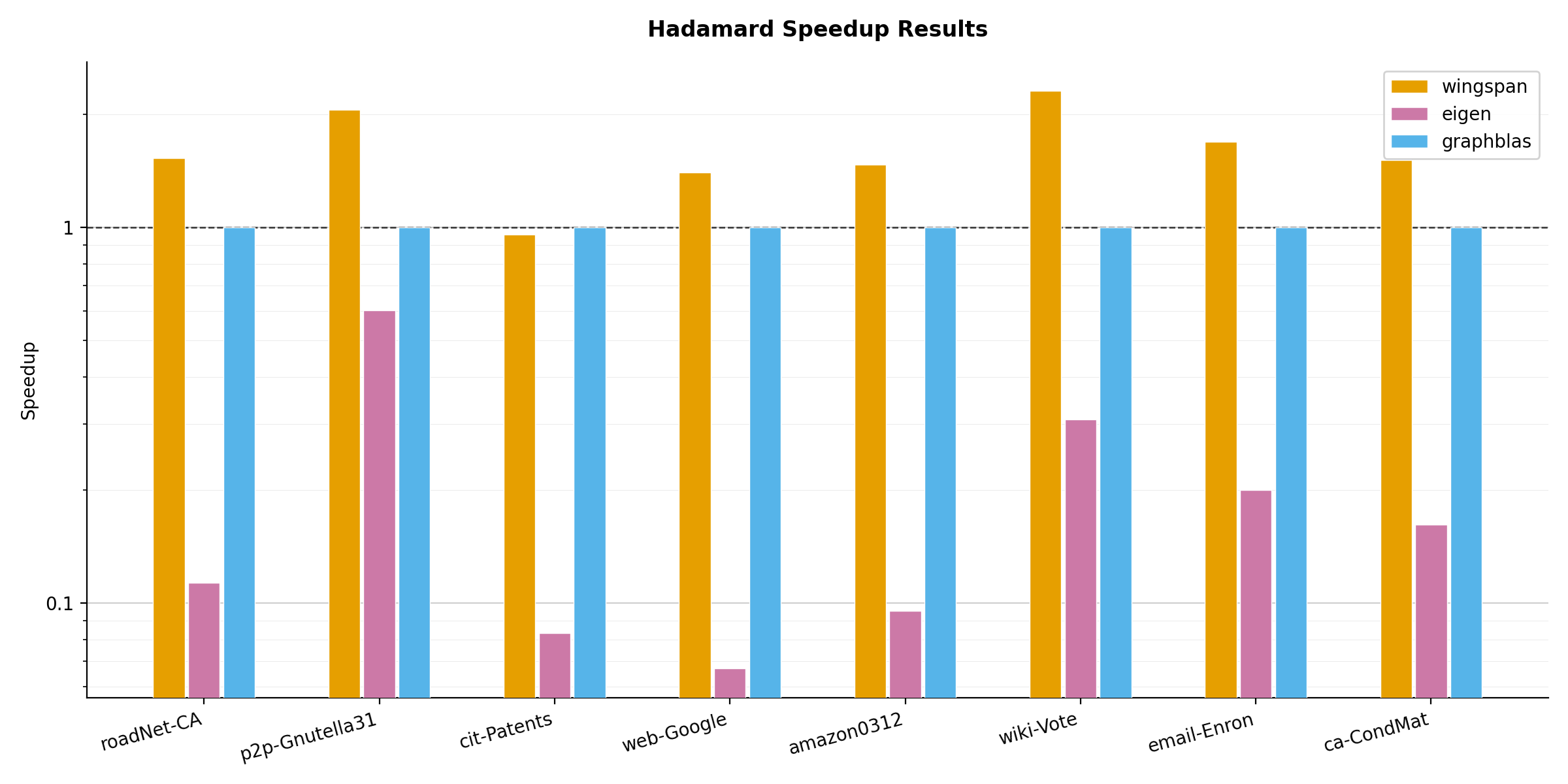}
    \end{subfigure}
    \caption{SpAdd and Hadamard speedups on the large sparse Zhang SNAP matrices, computing $C = A \text{ op } B$ where op $\in \{+, \odot\}$. Top row: identity case ($A = B$); bottom row: standard case ($B$ = random permutation of $A$).}
    \label{fig:spadd-hadamard}
\end{figure*}

Figure \ref{fig:spadd-hadamard} shows that WingSpan matches or outperforms the hand-optimized kernels across all four benchmarks. We compare SpAdd against MKL ($4.22\times$ speedup), GraphBLAS ($6.50\times$), and Eigen ($7.48\times$). We exclude MKL on Hadamard because it does not have a sparse Hadamard product. 

On the Hadamard benchmark, WingSpan performs comparably to GraphBLAS ($1.57\times$ speedup) on the standard dataset while being significantly better ($7.31\times$) on the identity dataset. GraphBLAS uses an extraction step for element-wise operations which counts the number of entries in each vector of the output without actually performing computation \cite{aznaveh2020parallel}. This preprocessing is wasted in the identity case since the output sparsity does not change. WingSpan's strategy does not require sparsity estimation, making it resilient to pathological inputs and reducing passes over the data.

Beyond element-wise operations, we consider sparse matrix sparse vector multiply. This kernel requires a sparse bytemap as the root element of the tensor tree, which will be accessed multiple times at the same index as $A$'s columns are accumulated. SpMSpV thus motivates the two uses of the merge level: enabling non-dense root levels and protecting against node and sibling dependencies. Wrapping the bytemap output in a merge level is sufficient to ensure safety.

\begin{figure}
    \centering
    \includegraphics[width=\columnwidth]{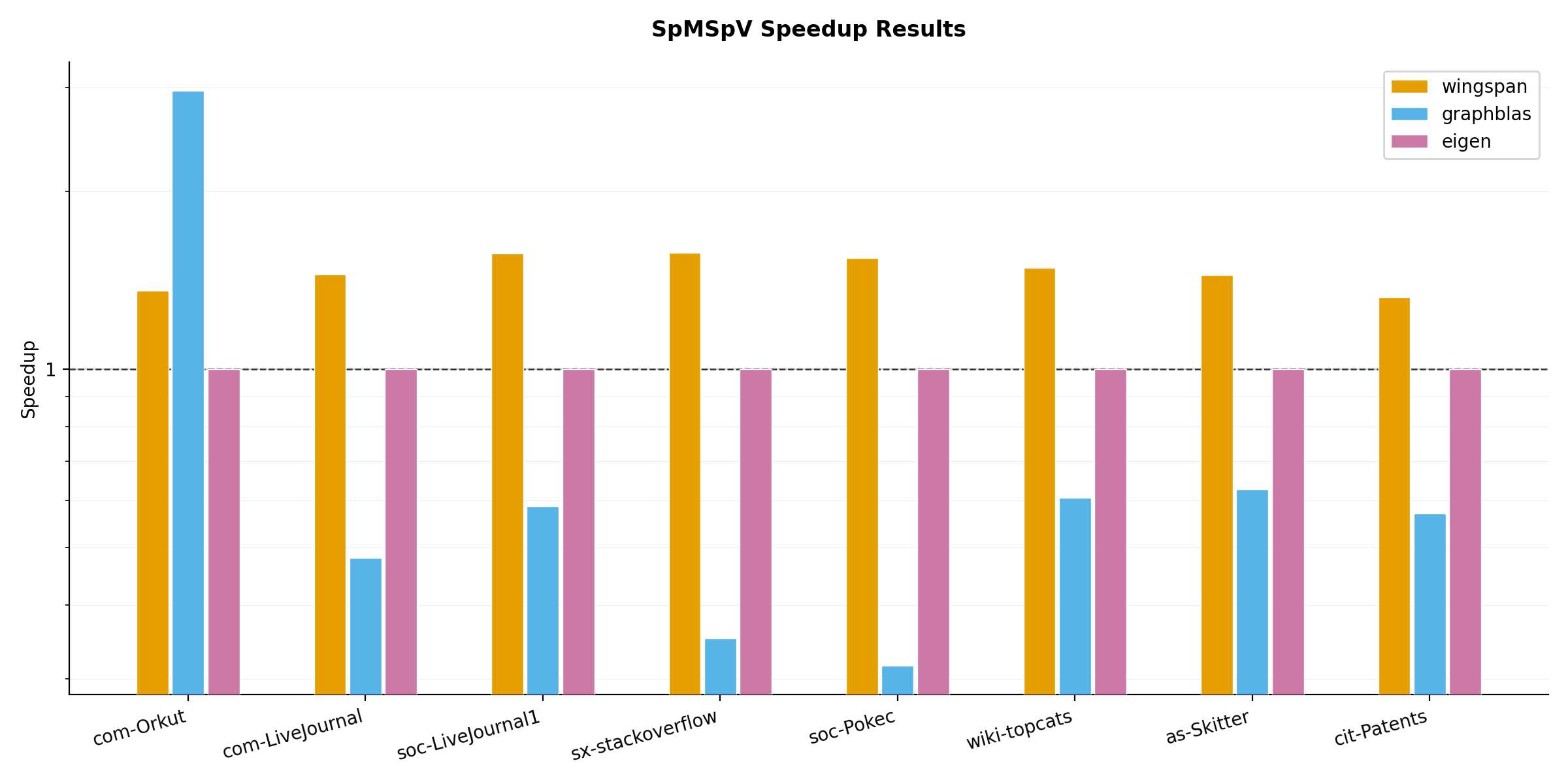}
    \caption{SpMSpV speedup results. We compute $y = A \cdot x$, where $x$ is a random sparse vector with 90\% sparsity.}
    \label{fig:spmspv}
\end{figure}

We evaluate on the 8 largest matrices, (under 1B non-zeroes), in the SuiteSparse SNAP collection \cite{Kolodziej2019}. Figure \ref{fig:spmspv} shows that WingSpan competes with Eigen ($1.47\times$ speedup) and GraphBLAS ($2.39\times$). The coalescing step has negligible overhead for large outputs. GraphBLAS outperforms on com-Orkut because this matrix has $3.4\times$ as many non-zeroes as the next largest matrix but has similar dimension sizes. WingSpan's sparse bytemap does not handle higher-density outputs as efficiently as GraphBLAS.

Finally, we benchmark the Matricized Tensor Times Khatri-Rao Product (MTTKRP) to evaluate WingSpan's performance on higher dimensional tensor algebra. MTTKRP computes $A_{ij}=\sum_{k,l}B_{ikl}C_{kj}D_{lj}$. We evaluate both dense MTTKRP, in which $A, C, D$ are all dense matrices, and sparse MTTKRP, where every matrix is sparse. In dense MTTKRP, no modifier levels are necessary since dense levels already support every form of parallel access. In the sparse case, we again use a sharded sparse list structure.

\begin{figure}
    \centering
    \begin{minipage}{0.5\linewidth}
        \includegraphics[width=\linewidth]{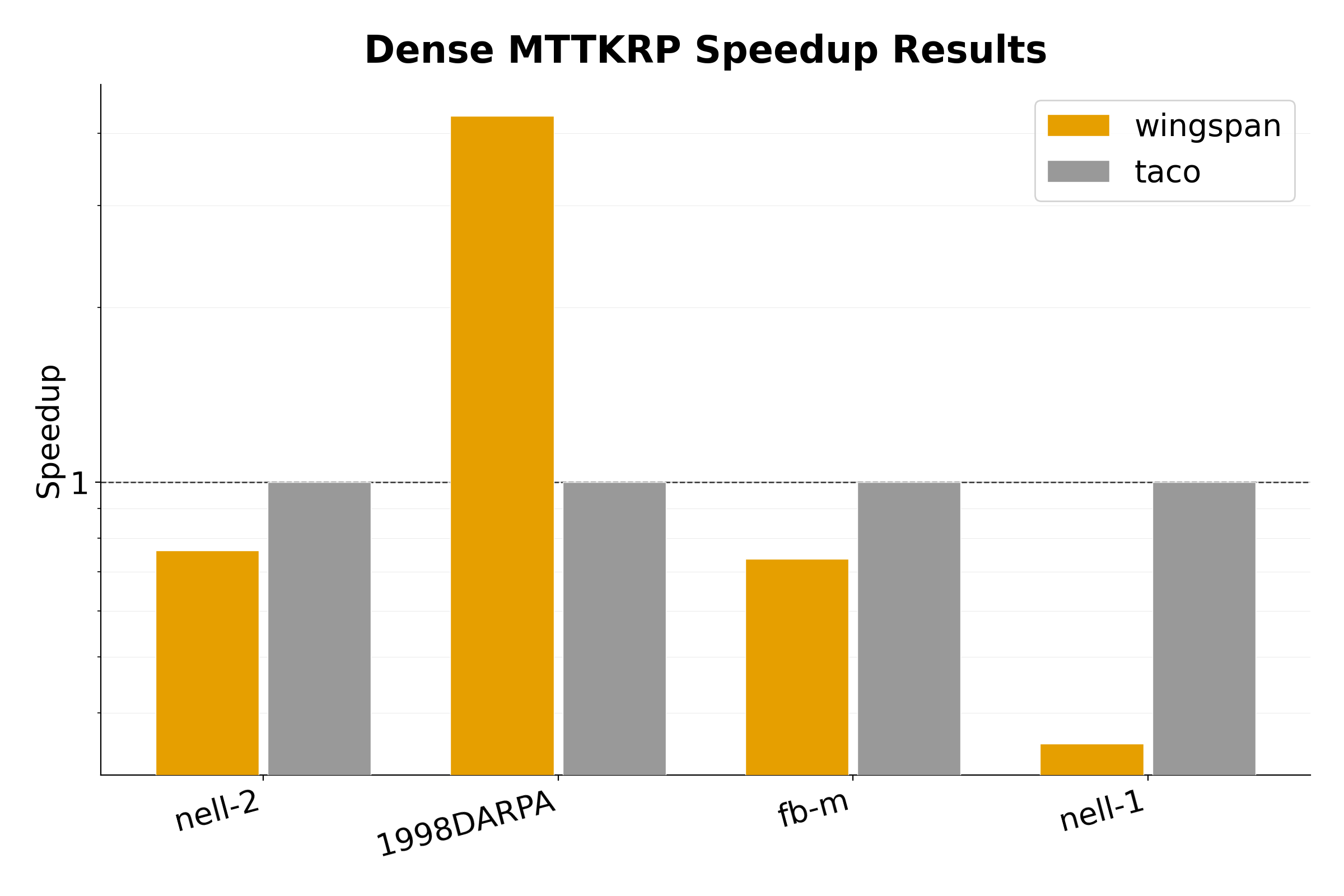}
    \end{minipage}%
    \begin{minipage}{0.5\linewidth}
        \includegraphics[width=\linewidth]{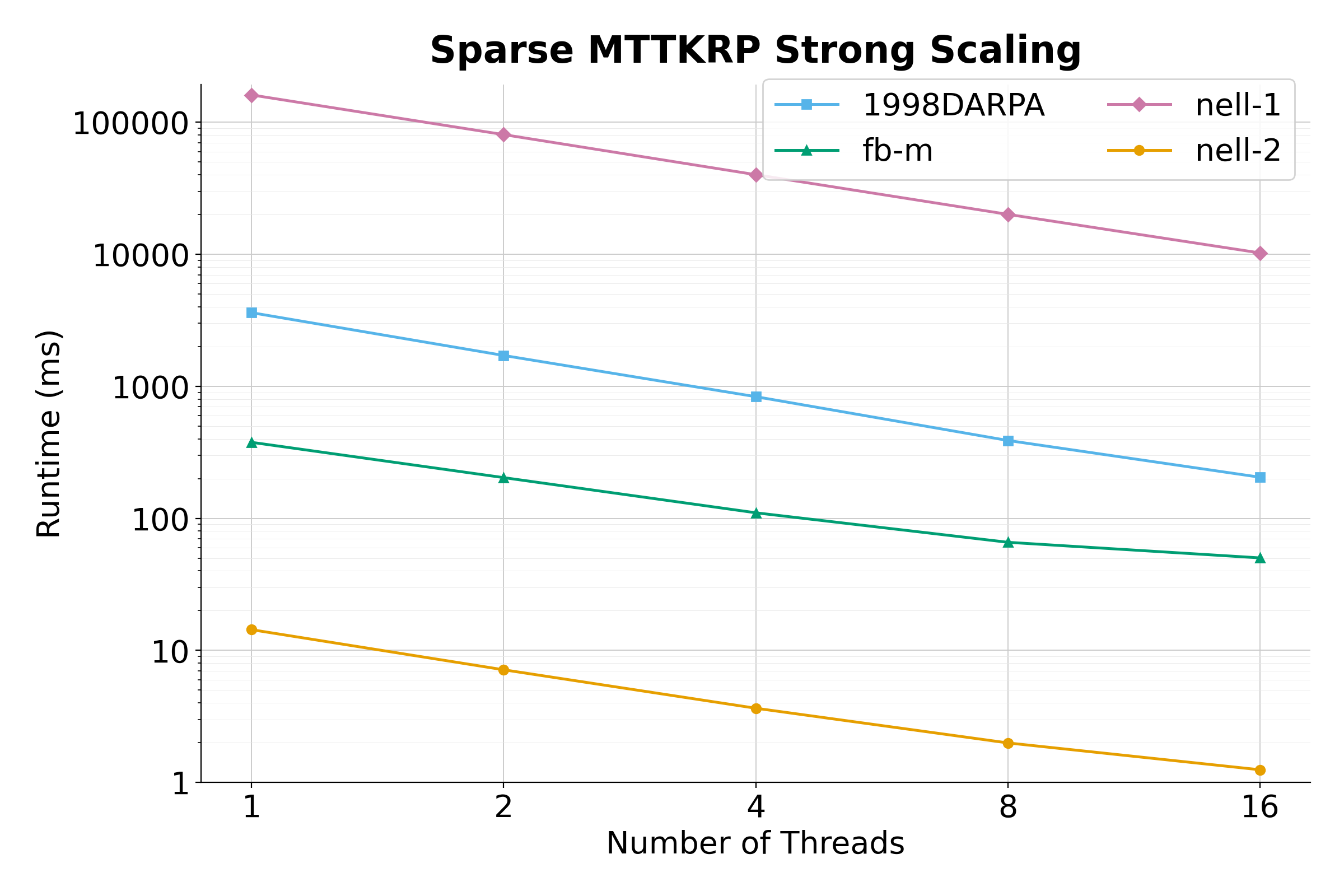}
    \end{minipage}

    \caption{Left: WingSpan's performance on MTTKRP relative to TACO. Right: sparse MTTKRP scaling.}
    \label{fig:mttkrp_speedup}
\end{figure}

We compare dense MTTKRP against TACO, which can parallelize dense output. The libraries lack an MTTKRP routine and are not included. Our selected competitors cannot parallelize the combination of sparse outputs and high-dimensional inputs, so we present strong scaling results for sparse MTTKRP. We evaluate using the three-mode tensors (under 1B non-zeroes) from the FROSTT collection as $B$ \cite{frosttdataset}. Both the dense and sparse kernels randomly generate $C$ and $D$. $C$ and $D$ are 99\% zero in the sparse version. Figure \ref{fig:mttkrp_speedup} shows both dense and sparse MTTKRP performance.

On average, WingSpan matches TACO's performance ($0.96\times$ speedup). TACO notably outperforms WingSpan by a factor of $2.82\times$ on the Nell-1 data because WingSpan's Finch backend is not optimized for this kernel. TACO is $3.83\times$ faster than Finch in serial on the same data, suggesting that WingSpan's parallelization strategy scales to 16 threads comparably with TACO's. The strong scaling results show that WingSpan effectively parallelizes across all available cores.

\subsection{Structured Tensor Algebra}

Finch supports a plethora of level formats, some of which represent additional patterns beyond simple sparsity. For example, \texttt{RunList} levels compress runs of repeated values together. Combining this pattern with sparsity represents images with monocolor backgrounds particularly well, as images tend to have many repeated nearby pixel values and the background can be treated as fill. To show how WingSpan's framework applies across kernels using arbitrary level formats, we compute an RGB histogram of 5 of the sparsest, (most fill background pixels), images from a Kaggle dataset of web screenshots \cite{aydos2020}. The input is an image represented with run-length encoding. The output stores each RGB channel as a hashmap.  We wrap the entire output in a merge level because any thread can access any RGB triple. To show our system can parallelize over structured output, we additionally replicate the SpAdd experiment with sparse run-length encoded images and outputs.

\begin{figure}
    \centering
    \begin{minipage}{0.49\linewidth}
        \includegraphics[width=\linewidth]{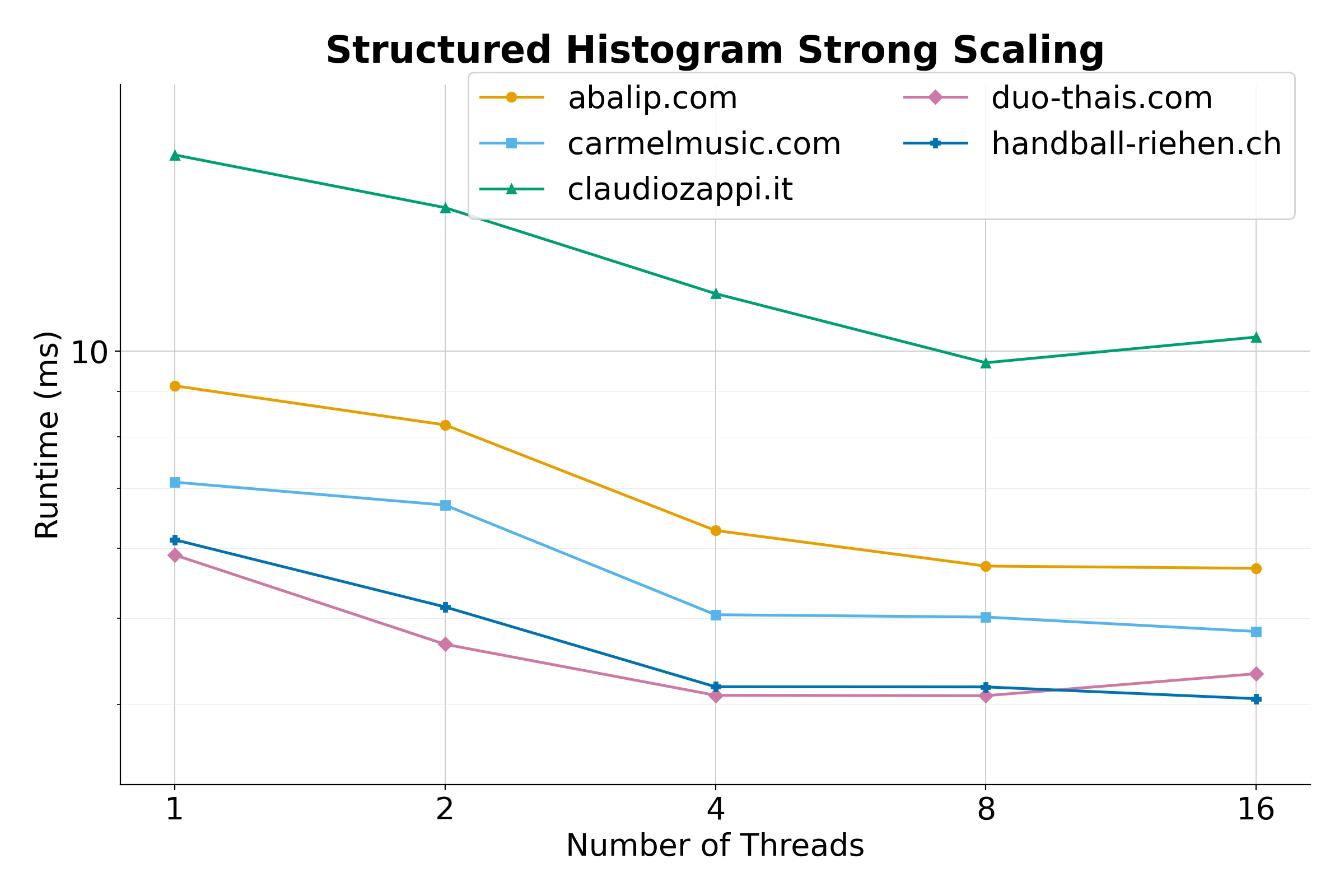}
    \end{minipage}
    \begin{minipage}{0.49\linewidth}
        \includegraphics[width=\linewidth]{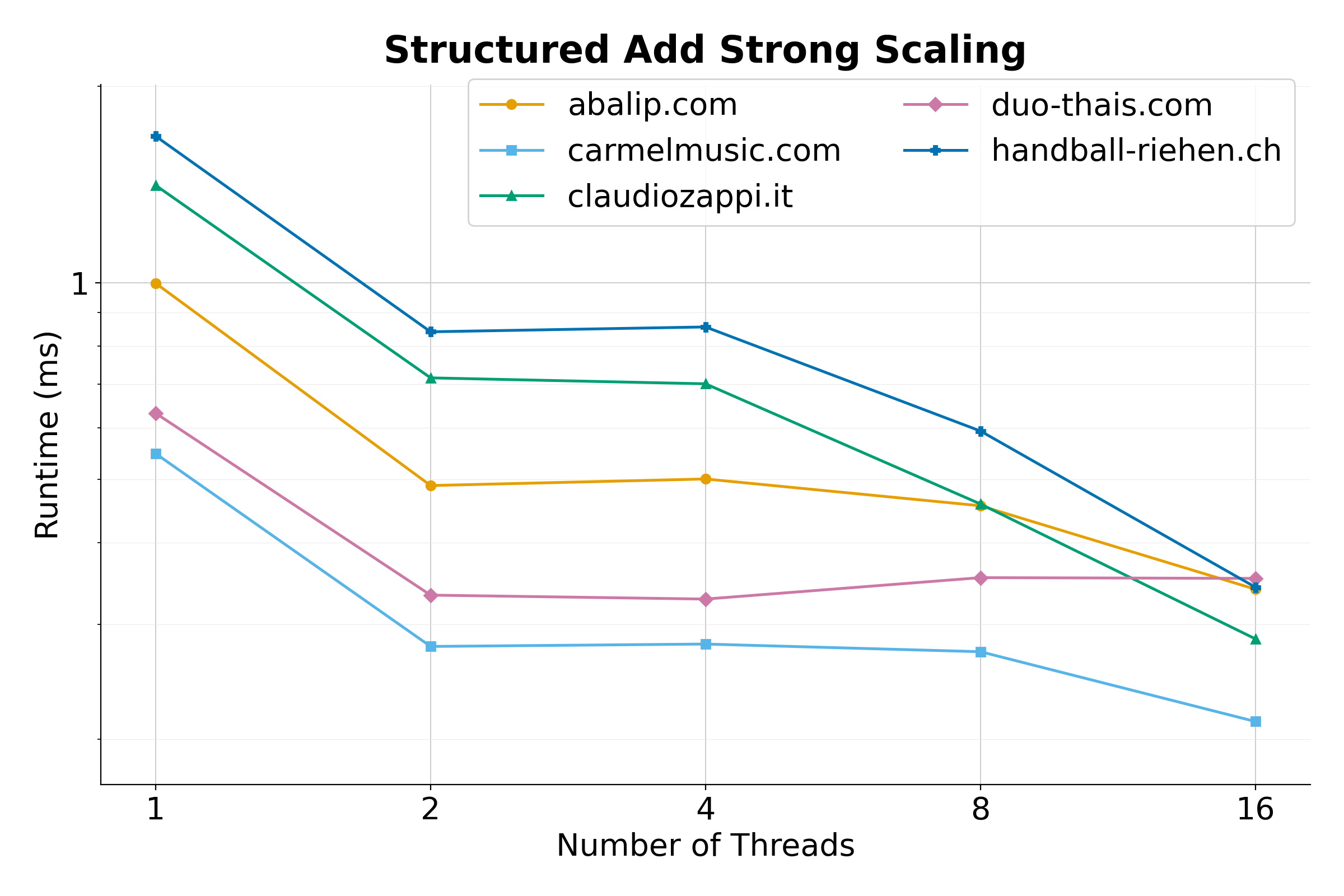}
    \end{minipage}
    \caption{Left: Histogram scaling. Indexing the output tensor with an RGB triple returns the number of pixels with that color. Right: Structured add scaling.}
    \label{fig:structured}
\end{figure}

Figure \ref{fig:structured} shows that WingSpan scales across both experiments and all five images. The input images are all small, making the overhead of increasing the threads sometimes mask the performance gain. However, every image performs better with more than one thread, and even in the worst cases, the effect of parallel overhead on runtime is minimal.

\section{Related Work}

The ecosystem is filled with sparse array programming frameworks, from libraries to compilers. Recent research on the compiler side has identified the need to support parallelism to maximize performance, but these techniques are not yet generalizable enough for widespread use. In this section, we review the landscape of sparse array frameworks.

\subsection{Sparse Libraries}
A wide variety of library routines, such as PETSc~\cite{balay_petsc_2020}, Armadillo~\cite{rumengan_pyarmadillo_2021}, OSKI~\cite{vuduc_oski:_2005}, MKL~\cite{noauthor_developer_2024}, CuSparse \cite{noauthor_cusparse_2024}, Eigen~\cite{guennebaud_eigen_2010}, GraphBLAS~\cite{kepner_mathematical_2016}, and LAGraph~\cite{mattson_lagraph_2019}, can handle some subset of parallel sparse kernels. However, library routines must be hand-written by experts for each specialized combination of format, operation, and architecture to get performance, presenting a significant implementation burden. As shown by our results section, the performance of library routines and support for parallelism varies significantly across kernels.

\subsection{Sparse Compilers}
Sparse tensor compilers differ from libraries in their ability to automatically generate optimized code for parameterizable formats and operations. This flexibility means they can support a much wider variety of
formats and can apply parallelization rules to a sparse kernel. However, current sparse tensor compilers lack the ability to emit code for a \textit{general} parallel sparse kernel, either leaving parallelism up to future work or requiring restrictive assumptions on tensor formats or intermediate processing.

Emblematic of this class of systems is the TACO compiler, which pioneered the sparse tensor compiler approach \cite{kjolstad_taco_2017,chou_format_2018}. TACO implements parallelism by annotating its generated code with OpenMP pragmas. However, it is limited to annotating outer loops and cannot emit code that handles scattering accumulation behavior. TACO is also restricted to strictly dense outputs for all kernels. 

Similar systems adjust the mechanism of parallel compilation, but have comparable constraints. For example, COMET represents parallelism as a pool of asynchronous tasks \cite{mutlu2020comet, tian_high-performance_2021}. However, its compilation strategy is highly specialized to sparse-dense kernels and cannot handle the sparse-sparse case. This restriction also applies to the output, which must be dense in all dimensions. Other sparse compilers such as SparseTIR \cite{ye_sparsetir_2023}, Taichi \cite{hu_taichi_2019}, and UniSPARSE \cite{liu_unisparse_2024} are comparably limited, while SDQL \cite{shaikhha_functional_2022} has left parallelism to future work. MLIR Sparse can parallelize, but performs all operations on a COO intermediate and then has to sort and sum this workspace to form the final result \cite{bik_compiler_2022}. NACHO, a concurrent work, introduces a general method for load-balancing sparse loop nests, which includes parallel writes to sparse outputs. However, it does not support workspaces and supports scatters indirectly with expand-sort-compress~\cite{chougule2026partitioningunstructuredsparsetensor}.

Few systems consider dependence analysis for sparse data structures. The sparse polyhedral framework is the closest related work \cite{strout_sparse_2018,zhao_polyhedral_2022, popoola_code_2023,strout_set_2012,strout_approach_2016}, extending polyhedral analysis to handle indirection arrays used in sparse formats and low-level sparse kernel code. While the previously described frameworks generate low-level sparse code from high-level dense representations, the sparse polyhedral framework focuses on optimizing already-sparse code. Sparse polyhedral approaches do not lend themselves readily to the productive fibertree-format abstractions of recent popularity.

Beyond the technical details of implementation, all of these frameworks treat parallelism as a feature but not a first-class citizen. No existing system has put forth a formal theory of fibertree dependencies. No system provides language features to enable expression of arbitrary parallel patterns, such as nested parallel loops or sparse local workspaces.

\section{Conclusion}

WingSpan is a fully parallel sparse tensor compiler and accompanying dependence theory for parallel fibertrees. Its hierarchical descriptor language expresses arbitrary parallel programming patterns across sparse and structured representations. Its dependence theory reasons about race conditions on fibertree structures, facilitating compiler-automated parallelism for sparse and structured tensor code. WingSpan's efficient and general language enables accelerated high-performance sparse applications. 

\section{Acknowledgements}

The authors would like to thank Thea Collin for many helpful discussions in formalizing early versions of the theory.

\bibliographystyle{ACM-Reference-Format}

\end{document}